\documentclass{article}
\usepackage[top=80pt,bottom=80pt,left=100pt,right=100pt]{geometry}
\usepackage[hidelinks]{hyperref} 
\usepackage{graphicx}
\usepackage{subcaption}
\usepackage{amsmath}
\usepackage{amssymb} 
\usepackage{bm}

\usepackage{tikz}        
\usetikzlibrary{shapes.geometric, arrows}
\usetikzlibrary{positioning}
\tikzstyle{rect} = [rectangle, draw, minimum width=2.5cm, minimum height=0.2cm]
\tikzstyle{elps} = [ellipse, draw, minimum width=2.5cm, minimum height=0.2cm]
\tikzstyle{arrow} = [thick,->,>=stealth]

\usepackage{booktabs}  
\usepackage{siunitx}   
\usepackage{algorithm}  
\usepackage{algpseudocode}
\usepackage{changepage} 
\usepackage{float}
\usepackage{ulem}  

\title{Scrap Composition Estimation in EAF and BOF: State-Space Models, Hyperparameters, and Validation
\footnote{
The research leading to these results has been performed within the CAESAR project (\href{https://caesarproject.eu/}{https://caesarproject.eu/}) and received funding from the European Commission’s Horizon Europe Programme under grant agreement n° 101058520. }
}

\author{Yiqing Zhou, Karsten Naert and Dirk Nuyens}
\date{April 2026}

\begin{document}

\maketitle

\abstract{
Accurate knowledge of scrap composition can increase the usage of recycled material to produce steel, reducing the need for raw ore extraction and minimizing environmental impact by conserving natural resources and lowering carbon emissions.
First, we introduce two state-space models for the elemental composition of scrap in Electric Arc Furnaces (EAF) and Basic Oxygen Furnaces (BOF): a linear model for elements that transfer entirely into steel, and a non-linear model for elements that partition between steel and slag. The models are fitted with the Kalman filter and the unscented Kalman filter, respectively, using only data already collected in the standard steel production process.  Crucially, the resulting scrap composition estimates can in turn be used to predict the elemental composition of future steel production.
Second, we analyze how key hyperparameters affect estimation accuracy and stability, and we provide practical guidelines for tuning them from expert knowledge and historical data.
Third, we validate the models on real BOF data from ArcelorMittal, using Cu and Cr as representative elements. 
Both filters outperform windowed non-negative least squares regression, a strong baseline method for scrap composition estimation, yielding reliable real-time estimates of scrap composition.
}

\section{Introduction}

Steel production forms the backbone of modern industry, underpinning infrastructure, transportation, and numerous other applications. In the face of growing environmental concerns and resource scarcity, the integration of recycled scrap steel into the manufacturing processes has become a vital step toward sustainability. By reducing reliance on raw materials, scrap steel conserves resources and significantly lowers the carbon footprint of steel production.

The Basic Oxygen Furnace (BOF) and Electric Arc Furnace (EAF) processes play crucial roles in utilizing recycled scrap steel. By maximizing scrap usage, it is possible to reduce carbon emissions and support the circular economy by promoting resource efficiency \cite{echterhof2021review, naidu2020basic, arzpeyma2021model}. However, variability and uncertainty in scrap steel composition present substantial challenges, affecting product quality, process modeling, and optimization \cite{de2019basic, hay2021review}.

Overcoming these challenges requires advanced modeling techniques capable of addressing variations in scrap composition, inaccuracies in weighing, and the distribution of elements during the melting process \cite{lahdelma1998amro, arzpeyma2021model}. Existing studies have utilized different methods such as ordinary and recursive least squares~\cite{sandberg2007scrap, birat2002quality}, maximum likelihood estimation~\cite{arzpeyma2021model}, sampling methods~\cite{gauffin29alloy}, and stochastic optimization~\cite{gaustad2007modeling}. While these approaches have succeeded in estimating element concentrations and, in some cases, their uncertainties, further advancements are needed to improve estimation accuracy and account for time-variant properties of scrap composition.

Building on these efforts, this study proposes a novel framework utilizing state-space models to reconstruct the composition of scrap steel in EAF and BOF processes. Leveraging production data,  Kalman and unscented Kalman filters are employed to provide real-time estimates of elemental fractions in scrap. By integrating linear and non-linear state-space models, the framework captures distinct element transfer behaviors between scrap, slag, and steel. 
Knowledge of the time-varying properties and the uncertainties allows to enhance process control, improve product quality, and supports the sustainability goals of the steel industry.

The remainder of this paper is organized as follows: Section~\ref{s:model} introduces the proposed models for describing element fraction changes in scrap and the mass balance equations for EAF and BOF processes. Section~\ref{s:kalman} details the modified Kalman filter and unscented Kalman filter algorithms used to fit these models, along with numerical results demonstrating the influence of hyperparameters and offering practical suggestions for their selection. Section~\ref{s:realdata} presents the application of the models to real BOF process data. Finally, Section~\ref{s:conclusion} concludes with key insights and directions for future research.

\section{Models and generation of data}\label{s:model}

The goal of this article is to develop models that describe the composition of scrap used in BOF or EAF processes in real time, propose methods to approximate these models for estimating scrap composition, and evaluate the validity of the models and their corresponding methods.
This section focuses on developing models that reflect the realistic behavior of scrap composition in actual processes. These models can then be used to generate synthetic data to test the algorithms used for reconstructing the model parameters. To this end, the section begins with the formulation of a theoretical model to describe scrap composition, 
comprising two key equations, the state update equation and the observation equation. The former one is detailed in Section~\ref{ss:state_equation}, while the latter one  is presented in Section~\ref{ss:observation_equation}.
The model is then utilized to generate synthetic data, which serves as a benchmark to assess the accuracy of the methods introduced in Section~\ref{s:kalman} for reconstructing the relevant parameters. The specifics of the data generation process are explained in Section~\ref{ss:generate_data}.

The models we put forward will follow the general structure of a state space model~\cite{durbin2012time}. We adopt the same notation as in~\cite{durbin2012time}. In particular, we denote the internal state at each timestep by \( \vec \alpha_t \). In its simplest form, we can think of this state as a statistical distribution representing the current composition of the scrap. However, it can also be expanded to include additional information, such as the state of the installation (whether it is BOF or EAF). This internal state is updated at each timestep based on a  \emph{state update equation}.

Additionally, at each timestep we consider a noisy measurement. An \emph{observation equation} will describe the measurement as a function of the state at that time.
Specifically, to model the fraction of a certain element X, we consider a model of the following form, where the notation is consistent with~\cite{durbin2012time},
\begin{align}
    y_t &= Z_t (\vec \alpha_t) + \varepsilon_t,  \quad  &\varepsilon_t &\sim \mathcal{N}(0, H_t),
    \label{eq:eq_obs}\\
    \vec \alpha_{t+1} &= T_t (\vec \alpha_t) + R_t (\vec \alpha_t) \, \vec \eta_t, \quad   &\vec \eta_t &\sim \mathcal{D}(\vec q, Q),
    \label{eq:eq_state}
\end{align}
where $t = 1, 2,\ldots$ and
\begin{itemize}
    \item \( y_t \) represents the observed measurement at time \( t \), 
    \item \( Z_t (\vec \alpha_t) \) is a function mapping the internal state \( \vec \alpha_t \) to the observed measurement,
    \item \( \varepsilon_t \) is the observation noise, assumed to follow a normal distribution $\mathcal{N}(0, H_t)$ with mean \( 0 \) and covariance \( H_t \),
    \item \( \vec \alpha_{t} \) is the state variable at time \( t\), representing the fraction of element X in each scrap type and  possibly additional parameters for partition coefficients, 
    \item \( T_t (\vec \alpha_t) \) is the state transition function, 
    \item \( R_t (\vec \alpha_t) \) is a function scaling the process noise,
    \item \( \vec \eta_t \) is the process noise, assumed to follow a distribution $\mathcal{D}(\vec q, Q)$ with mean \( \vec q \) and covariance \( Q \).
\end{itemize}
It is important to note that we do not require \( \vec \eta_t \) to have a zero mean or follow a normal distribution. This is because in our scenario, the state variable mainly represents the fraction of elements in scrap, and the values lie within the range \([0, 1]\), which cannot be guaranteed by a normal distribution. 
We will now explain each of these equations in more detail.

\subsection{The state and state update equation}
\label{ss:state_equation}
To motivate the state  update equation, which we will introduce in~\eqref{eq:state_general}, we can begin by first considering a standard random walk model, starting from $\vec \alpha_1^{\,\rm rw}$,
\begin{align}
    \vec \alpha_{t+1}^{\,\rm rw} = \vec \alpha_t^{\,\rm rw} + \mathcal N(\vec 0, \Sigma), \quad t = 1,2,\ldots,
    \label{eq:random_walk}
\end{align}
where $\Sigma$ is the covariance matrix of the new random step. This model is discussed in~\cite[Section~3.6.1]{durbin2012time}. This is a standard choice, however, because of the normal distribution, it has a significant drawback for our application: \( \vec \alpha_{t+1}^{\,\rm rw} \) can take on arbitrarily high or low values, which is unphysical when representing fractions. 
Moreover, since $\vec \alpha_{t+1}^{\, \rm rw}$ is a random variable, the mean and variance of $\vec \alpha_t^{\,\rm rw}$ can be written as 
\begin{align*}
    \mathbb{E}[\vec \alpha_{t+1}^{\,\rm rw}] 
    &= \mathbb{E}[\vec \alpha_{1}^{\,\rm rw}] , \\
     {\rm Var}[\vec \alpha_{t+1}^{\,\rm rw}] 
    &= {\rm Var}[\vec \alpha_{1}^{\,\rm rw}] + t \,\Sigma,
\end{align*}
where for $t\rightarrow\infty$, the expectation of $\vec \alpha_{t+1}^{\,\rm rw}$ does not change, while  the variance goes to infinity, which means that the uncertainty goes to infinity and the value of $\vec \alpha_{t+1}^{\,\rm rw}$ could be anywhere.

A more appropriate way to model the evolution of the state is through a convex combination of the form
\begin{align}
    \vec \alpha_{t+1} &= (1-\gamma)\,\vec\alpha_t + \gamma\,\vec \eta_t,
    \quad t = 1,2,\ldots,
    \label{eq:state_general}
\end{align}
where \( \gamma \) is a number in the interval \( [0,1] \),  and \( \vec \eta_t \sim \mathcal{D}(\vec q, Q) \) represents any distribution with mean \( \vec q \) and variance \( Q \), but constrained to the interval \( [0, 1] \). In this case, provided that the stochastic distribution \( \vec \alpha_1 \)   of the initial state is supported within the interval \( [0, 1] \), all subsequent \( \vec \alpha_t \) will also remain within this range.

Compared with the mean and variance of $\vec \alpha_t^{\,\rm rw}$ for the random walk in equation~\eqref{eq:random_walk},  since~\eqref{eq:state_general} is a convex combination, the mean and variance of \( \vec \alpha_{t+1} \) in equation~\eqref{eq:state_general} are given by
\begin{align}
    \vec a_{t+1} 
    &= \mathbb{E}[ \vec \alpha_{t+1}] 
    = (1-\gamma)^t \,\mathbb{E}[\vec \alpha_1] + (1 - (1 - \gamma)^t)\, \vec q
    , 
    \label{eq:a_tp1}\\
    P_{t+1} 
    &= {\rm Var}[\vec \alpha_{t+1}] 
    = (1 - \gamma)^{2t}\, {\rm Var}[\vec \alpha_1] + \frac{1 - (1-\gamma)^{2t}}{1 - (1 - \gamma)^2} \, \gamma^2 \, Q
    \label{eq:P_tp1}
    ,
\end{align}
and when $t$ increases the variance is now under control since
\begin{align}
    \vec a_\infty &= \lim_{t\rightarrow \infty} \mathbb{E}[ \vec \alpha_{t}] 
    = \vec q,
    \label{eq:lim_a_tp1}
    \\ 
    P_\infty &= \lim_{t\rightarrow \infty}{\rm Var}[\vec \alpha_{t}] 
     =  \frac{1 - ( 1 - \gamma)}{1 + ( 1- \gamma)} \, Q
    = \frac{\gamma}{2- \gamma} \, Q.
    \label{eq:lim_P_tp1}
\end{align}
It can be concluded that, the initial variance matrix \( P_1 \) becomes less important over time, and the variance of \( \vec \alpha_t \) will eventually stabilize at \( P_\infty \).

\subsection{The observation equation}
\label{ss:observation_equation}

\begin{figure}
    \centering
    \begin{tikzpicture}[node distance=0.7cm]
        \tikzset{every node/.style={font=\small}}  
        \node (scrap1) [rect] {Scrap $1$};
        \node (scrapd) [rect, below of=scrap1, draw=none] {$\vdots$};
        \node (scrapn) [rect, below of=scrapd] {Scrap $N_{\rm s}$};
        \node (hotmetal) [rect, below of=scrapn] {Hot metal};
        \node (add1) [rect, below of=hotmetal] {Addition 1};
        \node (addd) [rect, below of=add1, draw=none] {$\vdots$};
        \node (addn) [rect, below of=addd] {Addition $N_{\rm a}$};
        
        \node (bofeaf) [elps, right=1.5cm of hotmetal] {BOF or EAF};
        
        \node (slag) [rect, right=6cm of scrapn] {Slag};
        \node (steel) [rect, below of=slag] {Steel};
        \node (gases) [rect, below of=steel] {Gases};
    
        \draw [arrow] (scrap1.east) -- ++(0.7,0) |- (bofeaf.west);
        \draw [arrow] (scrapn.east) -- ++(0.7,0) |- (bofeaf.west);
        \draw [arrow] (hotmetal.east) -- ++(0.7,0) |- (bofeaf.west);
        \draw [arrow] (add1.east) -- ++(0.7,0) |- (bofeaf.west);
        \draw [arrow] (addn.east) -- ++(0.7,0) |- (bofeaf.west);
    
        \draw [arrow] (bofeaf.east) -- ++(0.5,0) |- (slag.west);
        \draw [arrow] (bofeaf.east) -- ++(0.5,0) |- (steel.west);
        \draw [arrow] (bofeaf.east) -- ++(0.5,0) |- (gases.west);
    
    \end{tikzpicture}
    \caption{\centering Schematic representation of the mass balance in steelmaking.}
    \label{fig:mass_balance}
\end{figure}
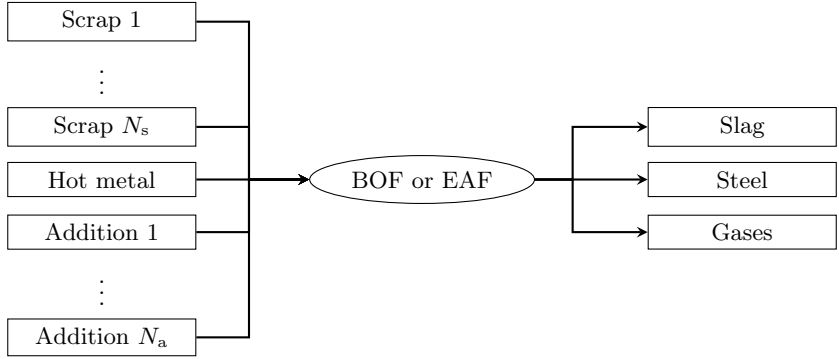

The observation equation is derived from the principle of mass balance within a single heat, where a \textit{heat} refers to a batch of metal processed in a furnace during one production cycle. 
A schematic representation of this process is provided in Figure~\ref{fig:mass_balance}.

The process involves reagents such as hot metal, various types of scrap, and additions, with the products being steel, slag, and gases. The mass balance equation for a chemical element $X$ at time $t$ is expressed as
\[
    m_{\text{hm}, X} + \sum_{s = 1}^{N_{\rm s}} m_{s, X}  + \sum_{a = 1}^{N_{\rm a}} m_{a, X}
    = m_{\text{steel}, X} + m_{\text{slag}, X} + m_{\text{gases}, X},
\]
where 
\begin{itemize}
    \item $m_{\text{hm}, X}$ is the mass of $X$ in hot metal,
    \item $N_{\rm s}$ and $N_{\rm a}$ are the number of distinct scrap types and addition types, respectively,
    \item $m_{\text{steel}, X}$, $m_{\text{slag}, X}$, and $m_{\text{gases}, X}$ represent the mass of $X$ in steel, slag, and gases.
\end{itemize}

For the chemical elements of interest in this work, we assume that their presence in gases and additions is negligible. However, we note that elements introduced via additions can, in principle, be easily incorporated into the model, as their mass and composition are typically known or controllable. In contrast, modeling elements that escape into the gases requires accurate measurement or estimation of gas mass. Expanding each of the masses of element $X$ into a product of the mass of this phase with the fraction of $X$ we obtain the following fundamental relation
\begin{equation}
    m_{\text{hm}} \, f_{\text{hm}, X} + \sum_{s = 1}^{N_{\rm s}} m_{s} \alpha_{s,X}
    = m_{\text{steel}} \, f_{\text{steel}, X} 
    + m_{\text{slag}} \, f_{\text{slag}, X},
    \label{eq:mass_balance}
\end{equation}
where
\begin{itemize}
    \item $f_{\text{hm}, X}$, $f_{\text{steel}, X}$ and $f_{\text{slag}, X}$ are the fraction of $X$ in hot metal, steel and slag, respectively,
    \item $\alpha_{s,X}$ is the fraction of $X$ in the $s$-th scrap type. It is denoted as $\alpha_{s,X}$ instead of $f_{s,X}$ because it also serves as the state variable $\vec \alpha_t$ in subsequent discussions.
\end{itemize}
We now make some remarks about the variables in~\eqref{eq:mass_balance}.
In practice, the masses of steel, slag, scrap and hot metal are known, either through direct measurement or, in the case of slag, estimated by an empirical model.
However, it is not possible to accurately measure or estimate these values. For example, during tapping from the BOF or EAF, steel and slag cannot be completely separated. As a result, only the mass of tapped steel can be measured, which is lower than the true steel mass. When generating synthetic data,  it is assumed that the uncertainty affects only the final observation, such as   the fraction of \( X \) in the steel, rather than the underlying process variables themselves.

Regarding the observation equation, the simplest modeling choice is to consider \( y_t \) as the measured mass of the chemical element $X$ in the steel, and \( Z_t(\vec \alpha_t) \) as a function describing this mass.  The relationship \( y_t = Z_t(\vec \alpha_t) \) should then hold. However, in practice,  observation noise is inevitable. Consequently, the observation equation is expressed as:
\[
y_t = Z_t (\vec \alpha_t) + \varepsilon_t, \quad \varepsilon_t \sim \mathcal{N}(0, H_t),
\]
where \( \varepsilon_t \) represents the observation noise, assumed to follow a normal distribution with zero mean and covariance $H_t$.

There are now two practical cases of interests, depending on whether the chemical element $X$ transfers to the slag or not. For example, Cu and Ni do not transfer to the slag, which makes the function $Z_t(\vec \alpha_t)$ a linear combination of $\vec \alpha_t$ and the model described by equations~\eqref{eq:eq_obs}--\eqref{eq:eq_state} reduces to a linear state space model. 
On the other hand, elements such as Cr and S do transfer to the slag and their masses in the slag cannot be measured directly and must be estimated. This estimation depends on the state variable $\vec \alpha_t$, which is the target of our inference. Consequently, the observation function $Z_t(\vec \alpha_t)$ becomes non-linear in $\vec \alpha_t$, complicating the model structure.
The details of these cases will be explained below.

\subsubsection{Linear state space model: e.g. Cu and Ni}
Chemical elements such as Cu and Ni do not form a significant amount of oxides~\cite{jung2010overview}, to the extend that we can neglect these elements in the slag in the mass balance \eqref{eq:mass_balance} by setting $m_{\text{slag}, X} = 0$. We obtain
\begin{equation}
    \sum_{s = 1}^{N_{\rm s}} m_s \, \alpha_{s, X}
    =m_{\text{steel}}\, f_{\text{steel}, X} - m_{\text{hm}}\, f_{\text{hm}, X}.
    \label{eq:mass_balance_cuni}
\end{equation}
The right-hand side is regarded as the observation \( y_t \), representing the mass of X from the scrap types, with some measurement noise. The measurements include $m_{\text{steel}}$, $f_{\text{steel}, X}$, $m_{\text{hm}}$, $f_{\text{hm}, X}$ and hence $y_t$ can be computed. The left-hand side is a linear combination of the state \( \vec \alpha_t \) and the input mass of the different scrap types at measurement \( t \). This leads to the observation equation
\begin{equation}
    y_t = Z_t (\vec \alpha_t) + \varepsilon_t
     \,= \vec m_t \cdot \vec \alpha_t + \varepsilon_t.
    \label{eq:obs_linear}
\end{equation}
where \( \vec m_t \) represents the input mass of scrap types for measurement \( t \), and $\vec \alpha_t$ represents the fraction of $X$ in the different scrap types for measurement $t$. Since this equation is linear, the model is a linear state space model, which can be written as:
\begin{align}
    y_t &= \vec m_t \cdot \vec \alpha_t + \varepsilon_t,  \quad  &\varepsilon_t &\sim \mathcal{N}(0, H_t),
    \label{eq:eq_obs_cuni}\\
    \vec \alpha_{t+1} &= (1 - \gamma) \, \vec \alpha_t + \gamma \, \vec \eta_t,  \quad  &\vec \eta_t &\sim \mathcal{D}(\vec q, Q).
    \label{eq:eq_state_cuni}
\end{align}

\subsubsection{Non-linear state space model: e.g. Cr and S}
Elements like Cr and S transfer from hot metal and scrap to steel and slag. We introduce a partition coefficient  \( \ell_X \), which is the ratio between the fraction of element $X$ in the slag and the fraction in the steel
\[  
\ell_X = \frac{f_{\text{slag}, X}}{f_{\text{steel}, X}}.
\]
Substituting \( \ell_X \) into the mass balance equation~\eqref{eq:mass_balance},  \( f_{\text{steel}, X} \) can then be written as
    \begin{equation}
        f_{\text{steel}, X} = 
        \frac{
            \displaystyle \sum_{s=1}^{N_{\rm s}}
             m_s \, \alpha_{s, X} + m_{\text{hm}}\, f_{\text{hm}, X}
        }{
            m_{\text{steel}} + m_{\text{slag}}\, \ell_X
        },
        \label{eq:mass_balance_crs_f}
    \end{equation}
or equivalently
    \begin{equation}
        m_{\text{steel}, X} = 
        \frac{
            \displaystyle \sum_{s=1}^{N_{\rm s}} m_s \, \alpha_{s, X} + m_{\text{hm}}\, f_{\text{hm}, X}
        }{
            1 + \ell_X \, m_{\text{slag}} / m_{\text{steel}}
        }.
        \label{eq:mass_balance_crs}
    \end{equation}
In  equation~\eqref{eq:mass_balance_crs}, the left-hand side can be interpreted as an observed variable measured with noise, while the right-hand side is a function of the state vector \( \vec \alpha_t \), which represents the fraction of element \( X \) in each scrap type. The key difference here, compared to elements that only go into steel, is the introduction of \( \ell_X \) and \( m_{\text{slag}} \).

A challenge in this model is determining an accurate estimate for \( \ell_X \) for each observation. One simple approach is to assume \( \ell_X \) is constant, but in reality, the partition coefficient depends on the overall state of the system, like the temperature and the input mass of reagents. We therefore allow some variability by assuming that \( \ell_X \) is a linear function of the fraction of iron oxide, denoted as $f_{\text{FeOn}, \text{slag}}$, which includes all types of iron oxides,
\begin{equation}
    \ell_{X} = c_1 + c_2 \, f_{\text{FeOn}, \text{slag}},
    \label{eq:linear_ell_cst}
\end{equation}
where \( c_1 \in \mathbb{R}\) and \( c_2 \in \mathbb{R} \) are parameters, depending on element $X$ and can be determined in two ways. The first option is to fix these values beforehand using either data or first principles. The second option is to treat them as variables that are adjusted during the model fitting process to optimize predictions for the scrap composition, leading to the form
\begin{equation}
    \ell_{X,t} = c_{1,t} + c_{2,t} \, f_{\text{FeOn}, \text{slag},t}.
    \label{eq:linear_ell_t}
\end{equation}
In this case, the state vector is extended to \( \vec \alpha_t^+ = [ \vec \alpha_t ,\, \vec c_t ]^\top\), where \( \vec c_t = [ c_{1,t} ,\, c_{2,t} ]^\top \).
It is important to note that the choice to approximate \( \ell_X \), either by~\eqref{eq:linear_ell_cst} or~\eqref{eq:linear_ell_t}, as a linear function of the iron oxide fraction in the slag, is a simple one. In reality, \( \ell_X \) could be a more complex, non-linear function of the iron oxide fraction, or could incorporate other factors such as the mass of slag or total input mass. 
However, the numerical results presented in Section~\ref{s:realdata} demonstrate that equation~\eqref{eq:linear_ell_t} performs effectively for the real data considered in this study.
Therefore, the number of components in \( \vec c_t \) is not strictly limited to two, as shown in this example.

Continuing with the second option, we now interpret the mass balance equation~\eqref{eq:mass_balance_crs} as:
\begin{equation}
    y_t =  Z_t (\vec \alpha_t^+) + \varepsilon_t, \quad \varepsilon_t \sim \mathcal{N}(0, H_t),
    \label{eq:obs_z_cr}
\end{equation}
where \( \vec \alpha_t^+ \) is the extended state vector, and \( Z_t \) is a non-linear, time-varying function that represents our estimate of the mass of element \( X \) in the steel, based on the right hand side of equation~\eqref{eq:mass_balance_crs}. The observation function now represents the mass of element \( X \) in the steel, and it is no longer a simple linear combination of components of \( \vec \alpha_t^+ \). Instead, it depends on the fraction of elements in the scrap types as well as additional parameters for the partition coefficients.
The observation function is
\begin{equation}
    Z_t (\vec \alpha^+_t) = Z_t (\vec \alpha_t, \vec c_t) = 
        \frac{
            \vec m_t \cdot \vec \alpha_{t} + m_{\text{hm}, t} \, f_{\text{hm}, t}
        }{
            1 + (c_{1,t} + c_{2,t}  f_{\text{FeOn, slag}, t}) \, m_{\text{slag}, t} / m_{\text{steel}, t} 
        },
    \label{eq:zt_crs}
\end{equation}
which is non-linear since $c_{1,t}$ and $c_{2,t}$ are part of $\vec \alpha_t^+$. 
The observation $y_t$ can be calculated by the multiplication of $m_{\rm steel}$ and $f_{{\rm steel}, X}$, which can be measured during steelmaking.

The state update equation remains the same as in the linear case~\eqref{eq:eq_state_cuni}.
Therefore, the model can now be expressed as a non-linear, non-Gaussian state space model
\begin{align}
    y_t 
    &= Z_t(\vec \alpha_t^+) + \varepsilon_t \notag \\
    &= 
        \frac{
            \vec m_t \cdot \vec \alpha_{t} + m_{\text{hm}, t} \, f_{\text{hm}, t}
        }{
            1 + (c_{1,t} + c_{2,t}  f_{\text{FeOn, slag}, t}) \, m_{\text{slag}, t} / m_{\text{steel}, t} 
        } + \varepsilon_t,
    \varepsilon_t 
    \sim \mathcal{N}(0, H_t),
    \label{eq:eq_obs_crs}\\
    \vec \alpha^+_{t+1} 
    &= \begin{bmatrix} \vec \alpha_{t+1} \\ \vec c_{t+1} \end{bmatrix} 
    = (1 - \gamma) \begin{bmatrix} \vec \alpha_t \\ \vec c_t \end{bmatrix}
    + \gamma \begin{bmatrix} \vec \eta_t \\ \vec \theta_t \end{bmatrix},
    \quad  
    \vec \eta_t 
     \sim \mathcal{D}(\vec q, Q), \quad \vec \theta_t \sim \mathcal{D}(\vec q_c, Q_c),
    \label{eq:eq_state_crs} 
\end{align}
where the random vector $\vec \theta_t$ behaves in a similar manner to $\vec \eta_t$ as the process noise for $\vec c_t$ and $\vec \theta_t \sim \mathcal{D}(\vec q_c,\, Q_c)$ represents any distribution with mean $\vec q_c$ and variance $Q_c$, and we have $\lim_{t \rightarrow \infty}\mathbb{E}[\vec c_t] = \vec q_c$.
It is essential to highlight that the state variable is now represented by $\vec{\alpha}^+_t$ instead of $\vec{\alpha}_t$, as was previously used in the linear state space model. 
For convenience, the distribution $\mathcal{D}(\vec q_c, Q_c)$ is chosen as a Beta distribution and the parameter $\gamma$ applies to both $\vec{\alpha}_t$ and $\vec{c}_t$ in the equation above. It is of course also possible to have a different value of $\gamma$ in the state update for \( \vec c_t \), but we will not pursue that here.

\subsection{Generation of data}
\label{ss:generate_data}
We aim to be able to generate realistic data to test model misspecification. Rather than generating all data from scratch, we incorporate real production data to ensure realistic input distributions and naming conventions. In particular, we use actual input masses and naming conventions for $N_{\rm s} = 45$ different scrap types over $T=20000$ heats. The scrap input masses for each heat are taken directly from real production data. The generated data are introduced below. 
The method is also applied to real data and the results will be discussed in Section~\ref{s:realdata}.

\subsubsection{Generation of  state variables: fraction of elements}
\label{sss:generate_state_frac}
In the linear state space model~\eqref{eq:eq_obs_cuni}--\eqref{eq:eq_state_cuni}, the state variable represents the fraction of elements in the scrap types. In the non-linear state space model~\eqref{eq:eq_obs_crs}--\eqref{eq:eq_state_crs}, the state variable represents both the fraction of elements in the scrap types and the parameters for the partition coefficients in~\eqref{eq:linear_ell_t}. In this subsection, we will  explain how to generate the fraction of elements. The parameters for the partition coefficients will be introduced in the next subsection.
It is worth noting that \( \vec{\alpha}_t \) is a random variable, representing a stochastic distribution. Therefore, ``generating data'' refers to creating a specific instance based on this stochastic distribution.

According to equations~\eqref{eq:eq_obs_cuni}--\eqref{eq:eq_state_crs}, which will be used to generate data, $\vec \alpha_t$ represents the element's fraction in the different scrap types.  The average of $\vec \alpha_t$ over a large enough time horizon will approach \( \lim_{t \rightarrow \infty} \mathbb{E}[\vec \alpha_t] = \vec q\), see~\eqref{eq:lim_a_tp1}.
Estimated mean values of elemental fractions for each scrap type, derived from real production data, are also used to define representative values for $\vec \alpha_1$ and $\vec q$. Examples of these values for selected elements are listed in Table~\ref{tab:mean_concentration}.

\begin{table}[htbp]
\caption{\centering The mean values  for the generated element fraction of scrap types (unit: ppm).}
\label{tab:mean_concentration} \centering
\begin{tabular}{SSSSS}\toprule
{scrap type } & {Cu}      & {Ni}      & {Cr}       & {S}       \\ \midrule
{$01$}        & {$208.80$}  & {$227.70$}  & {$180.00$}   & {$99.00$ }  \\ 
{$02$}        & {$900.00$}  & {$227.70$}  & {$900.00$}   & {$351.00$}  \\ 
{\vdots} & {\vdots}   & {\vdots}    & {\vdots}   & {\vdots}   \\ 
{$45$}        & {$3060.99$} & {$1147.95$} & {$1601.64$}  & {$342.00$}  \\ \bottomrule
\end{tabular}
\end{table}

Before talking about the values of $Q$ and $P_1$ for generating data, we will first elaborate on choosing the parameter $\gamma$. 
This parameter is chosen such that the half-life period of the state variable is 1000 heats, meaning
\begin{equation}
    \gamma = \frac{\ln 2}{1000}.
    \label{eq:gamma}
\end{equation}
As indicated by equation~\eqref{eq:lim_P_tp1}, the covariance matrix of \( \vec{\alpha}_t \) converges to \( P_\infty = (\gamma / (2 - \gamma)) \, Q \), independent of the initial value \( P_1 \), which bounds the interval for most probable values of $\vec \alpha_t$. The values of $P_\infty$ further depend on $\vec q$, as larger element fractions typically have higher uncertainty. Assuming the element fractions are uncorrelated across different scrap types, the covariance matrix $P_\infty$ is diagonal. We define its diagonal elements to be
\[
   (P_\infty)_{i,i} = (0.042 \, q_i)^2,\quad i = 1,2,\ldots, N_{\rm s}.
\]
This implies that, over large time horizons, the state variable will predominantly lie within the interval $\big[\, (1 - 0.042)\, \vec q, \, (1+0.042)\, \vec q \,\big]$. The value $0.042$ is chosen as a reasonable estimate for data generation.
From equation~\eqref{eq:lim_P_tp1}, the matrix $Q$ is then also diagonal and can be computed as $Q = ((2 - \gamma)/\gamma) \, P_\infty$. Consequently,
\[
Q_{i,i} = 5 \, q_i^2, \quad i = 1,2,\ldots, N_{\rm s}.
\]
Since $\gamma$ is a constant, the values of $Q$ and $P_\infty$ are proportional to the corresponding value in $\vec q$.
It also means that given the value of $\gamma$ and $Q$, we can calculate the value of $P_\infty$.

Since the fraction values must remain within the range \([0, 1]\), using a normal distribution for \( \mathcal{D}(\vec q, Q) \) would not be suitable, as it could generate values outside this interval. Therefore, we model the components of \( \vec \eta_t \) using a Beta distribution \( \mathcal{D}(q_i, Q_{i,i}) = \text{Beta}(u_i, v_i) \), where \( q_i \) is the mean and \( Q_{i,i} \) is the variance of the Beta distribution. If a random variable \( X  \sim \mathcal{D}(q_i, Q_{i,i}) = \text{Beta}(u_i, v_i) \), the parameters \( u_i \) and \( v_i \) can be computed as follows:

\begin{align*}
    u_i &= \frac{q_i^2 \, ( 1 - q_i)}{Q_{i,i}} - q_i,\qquad
    v_i  = \frac{u_i}{q_i} - u_i.
\end{align*}
Thus with the given mean vector \( \vec q \) and covariance matrix \( Q \), the values of \( u_i \) and \( v_i \) for each component can be calculated.

Using the initial state variable \( \vec \alpha_1 \),  \( \gamma \), and the generated random vector \( \vec \eta_t \), the fraction of elements in the scrap types can be generated using equation~\eqref{eq:eq_state_cuni}. This process generates the state variable for the linear state space model~\eqref{eq:eq_obs_cuni}--\eqref{eq:eq_state_cuni}, and also part of the state variable in the non-linear state space model~\eqref{eq:eq_obs_crs}--\eqref{eq:eq_state_crs}.

\subsubsection{Generate the state variables: partition coefficients}
In the non-linear state space model~\eqref{eq:eq_obs_crs}--\eqref{eq:eq_state_crs}, the state variable $\vec \alpha_t^+$ consists of two parts: the element fraction $\vec \alpha_t$ and the parameters for the partition coefficients $\vec c_t$. 
We follow the same procedure to generate $\vec \alpha_t$ for element Cr and S according to Section~\ref{sss:generate_state_frac} and Table~\ref{tab:mean_concentration}.
Following the same reasoning as for $\vec \alpha_1$,  we set $\vec c_1 = \vec q_c$.
Since according to equation~\eqref{eq:eq_state_crs},
\[
    \vec c_{t+1} = (1 - \gamma) \, \vec c_{t} + \gamma \, \vec \theta_t, \quad \vec \theta_t \sim \mathcal{D}(\vec q_c, Q_c),
\]
where the value of $\gamma$ has been set to $\ln 2 / 1000$ according to~\eqref{eq:gamma}. After setting the value for $\vec c_1$, $\vec q_c$ and $Q_c$, we can calculate the partition coefficient at the $t$-th heat using~\eqref{eq:linear_ell_t}
\[
    \ell_t = 
    c_{1,t} + c_{2,t} \, f_{{\rm FeOn, slag}, t}.
\]

It is assumed that the value of the partition coefficients for Cr here is always around $10$, which might be influenced by many other factors in real production~\cite{jung2010overview, ono1992partition, battle1989equilibrium}.  Thus we set \( \vec q_c =  [ 9.7 ,\, 0.01 ]^\top \).
Similar to~\eqref{eq:lim_P_tp1}, 
\[
    P_{c, \infty} := \lim_{t\to\infty} {\rm Var}[ \vec c_t] = \frac{\gamma}{2 - \gamma} Q_c,
\]
where $P_{c, \infty}$ is a diagonal matrix and the diagonal elements of $P_{c,\infty}$ are defined as: 
\[
    (P_{c, \infty})_{i,i} = \left( 0.01 \, q_{c,i} \right)^2,
\]
where $q_{c,i}$ is the $i$-th component of vector $\vec q_c$. Thus $Q_c$ can be calculated by
\[
    Q_c = \frac{2 - \gamma }{\gamma} P_{c, \infty}.
\]
The random vector $\vec \theta_t$ in~\eqref{eq:eq_state_crs} can now be generated by a normal distribution with mean $\vec q_c$ and covariance matrix $Q_c$.

\subsubsection{Generation of other parameters}
The observation equations~\eqref{eq:eq_obs_cuni} and~\eqref{eq:eq_obs_crs} are derived from the mass balance equations~\eqref{eq:mass_balance_cuni} and~\eqref{eq:mass_balance_crs} with added observation noise. To generate data for each observation, we need the following parameters: \({m}_s \), \( m_{\rm hm} \), \( f_{\rm hm} \), \( m_{\rm slag} \), \( m_{\rm steel} \) and \( f_{\rm FeOn, slag} \) at each time step $t$.

As mentioned at the beginning of Section~\ref{ss:generate_data}, the input mass of scrap of the $s$-th scrap type, \( m_s \), is not synthetically generated, but taken from real production data, reflecting actual scrap usage patterns during production. 
The parameters $m_{\rm hm}$, $f_{{\rm hm}, X}$, $m_{\rm slag}$, $m_{\rm steel}$ and $f_{\rm FeOn, slag}$ are also obtained from production records. Thus the steel composition $f_{\rm steel, X}$ can be calculated from the mass balance equation~\eqref{eq:mass_balance_cuni} or~\eqref{eq:mass_balance_crs}.

It is also important to remind the reader, as noted in Section~\ref{ss:observation_equation} following~\eqref{eq:mass_balance}, that in practice parameters \(  m_s \), \( m_{\rm hm} \), and \( f_{{\rm hm},X}\) are directly measured and used, while \( m_{\rm slag} \), \( m_{\rm steel} \), and \( f_{\rm FeOn, slag} \) are estimated based on other measurements and empirical models. All values include some level of noise. For simplicity,  we assume that these values are accurate and account for uncertainty only through the observation noise.

Observation noise is modeled differently for linear and non-linear state space models:
\begin{itemize}
    \item \textbf{Linear state space model.} For elements like Cu and Ni, which do not enter the slag, the observation variable represents the mass of the element for the scrap. Since the observation $y_t$ is calculated based on the right hand side of  equation~\eqref{eq:mass_balance_cuni}, the variance of the observation noise is
    \begin{equation}
    m_{\text{steel}}^2 \, {\rm Var}[f_{\text{steel}, X}]
    +
    m_{\text{hm}}^2 \, {\rm Var}[f_{\text{hm}, X}].
    \label{eq:estimate_Ht}
    \end{equation}
    For our data, for Cu and Ni, we have \( m_{\rm steel} \approx 330 \, [\text{t}] \) and \( m_{\rm hm} \approx 280 \, [\text{t}] \). The standard deviations of \( f_{\text{steel}, X} \) and \( f_{\text{hm}, X} \) are set to \( 12 \, [\text{ppm}] \) and \( 5 \, [\text{ppm}] \), respectively. Thus, the variance of the observation noise is \( 330^2 \times 12^2 + 280^2 \times 5^2 \), with unit \( [\text{g}^2] \).

    \item \textbf{Non-linear state space model.} For elements like Cr and S, which partition between slag and steel, the actual observation is the mass of the element in steel, given by the product \( m_{\rm steel} f_{{\rm steel}, X} \), which differs from the earlier definition of the observation variable \( y_t \) in linear model.  To be consistent and to avoid confusion, we model the variance of the observation noise as $m_{\rm steel}^2 \, {\rm Var}\left[ f_{\rm steel, X}\right]$.
    Based on practical experience, the variance of the observation noise is set to \(  330^2 \times 4^2  \), with unit $[ \rm g^2]$.
\end{itemize}

\section{Fitting the model}
\label{s:kalman}
In this section, we introduce two methods for fitting the models presented in the previous section. For the linear model, we employ the Kalman filter~\cite{durbin2012time, grewal2014kalman}, while for the non-linear model, the unscented Kalman filter (UKF)~\cite{durbin2012time, julier2004unscented, menegaz2015systematization, wan2000unscented} is utilized. Additionally, we analyze the influence of hyperparameters and provide practical recommendations for selecting these hyperparameters in applications.

\subsection{Algorithms: Kalman filter and unscented Kalman filter}
\label{ss:alg-kf-ukf}
\subsubsection{Linear state space model: Kalman filter}
In the previous section, we introduced the model denoted by equations~\eqref{eq:eq_obs_cuni}--\eqref{eq:eq_state_cuni} for elements like Cu.

These equations define a linear state space model, where the state variables are not observable, while the observations are given. Once the parameters and the initial state are defined, a standard method for fitting such models is the Kalman filter~\cite{durbin2012time}.

The Kalman filter is an algorithm used to estimate the state of a model from noisy observations. It is particularly useful in situations where you want to track a variable (such as the fraction of a certain element in scrap) over time, but the available data is noisy or uncertain. The filter works by combining predictions based on a system's model with actual observations, providing an estimate of the system's state. Additionally, the Kalman filter and its variations can also be applied to systems where the state variables are constrained within a certain domain, as demonstrated in our case and e.g.\,in~\cite{torn2006boundary}.

In our model, as shown in equation~\eqref{eq:eq_state_cuni}, \( \vec \eta_t \) follows a Beta distribution. However, since \( \vec \eta_t \) represents a small disturbance in the state variable, it is reasonable to approximate it using the Kalman filter which assumes that \( \vec \eta_t \) follows a normal distribution. For more details on the Kalman filter algorithm, see~\cite[Section~4.3]{kalman1960approach}.
Our equation~\eqref{eq:eq_state_cuni} differs slightly from~\cite[equation~(4.12)]{kalman1960approach}, where the mean of the process noise is assumed to be 0. In our case, we assume the mean is \( \vec q \). As a result, the prediction step in our Kalman filter will differ slightly from the one in~\cite{kalman1960approach}. For completeness, the full algorithm is provided in Algorithm~\ref{alg:km_cuni}.

For convenience, we present the linear Gaussian state space model based on equations~\eqref{eq:eq_obs}--\eqref{eq:eq_state}, which we will use to approximate the model in equations~\eqref{eq:eq_obs_cuni}--\eqref{eq:eq_state_cuni}:
\begin{align}
    y_t &= Z_t \, \vec \alpha_t + \varepsilon_t,  \quad  &\varepsilon_t &\sim \mathcal{N}(0, H_t),\\
    \vec \alpha_{t+1} &= T_t \, \vec \alpha_t + R_t \, \vec \eta_t, \quad   &\vec \eta_t &\sim \mathcal{N}(\vec q, Q), \quad t = 1, 2, \ldots \\
        &   &  \vec \alpha_1 &\sim \mathcal{N} (\vec a_1, P_1),\notag
\end{align}
where  \( Z_t = \vec m_t^\top \in \mathbb{R}^{1 \times N_{\rm s}} \) represents the input mass of scrap types. The state variable \( \vec \alpha_t \in \mathbb{R}^{N_{\rm s}} \) represents the fraction of elements in the scrap types, thus $Z_t \, \vec \alpha_t = \vec m_t \cdot \vec \alpha_t$ is the inner product. The covariance matrix of measurement noise is denoted as \( H_t \in \mathbb{R}^{1 \times 1} \). The matrices $T_t$ and $R_t$ are diagonal, featuring $(1 - \gamma)$ and $\gamma$ on their diagonals, respectively, which are determined by the correlation parameter of time steps. The process noise \( \vec \eta_t \) has mean \( \vec q \) and covariance matrix \( Q \).
The initial state $\vec \alpha_1$ is defined by the initial values \( \vec a_1 \) and \( P_1 \).  

\begin{algorithm}
\caption{Kalman\_Step (($\vec m_t, \, m_{\rm steel, t},\, f_{{\rm steel}, X, t},\, m_{{\rm hm}, t},\, f_{{\rm hm}, X, t}$); ($\vec a_t, P_t$); $\gamma$, $\vec q$, $Q$, $H_t$)}
\label{alg:km_cuni}
\begin{algorithmic}
\State  $y_t = m_{\rm steel, t} f_{{\rm steel}, X, t} - m_{{\rm hm}, t} f_{{\rm hm}, X, t}$ 

\State  $Z_t = \vec m_t^\top$ 

\vspace{0.5em}
\State \textbf{Update state variables based on new measurement $(\vec m_t, y_t)$}
\State \quad  $K_t = P_t Z_t^\top (Z_t P_t Z_t^\top + H_t)^{-1}$
\State \quad  $\vec a_{t|t} = \vec a_t + K_t (y_t - Z_t\vec a_t )$
\State \quad  $P_{t|t} = (I - K_t Z_t)P_t$

\vspace{0.5em}
\State \textbf{Predict the next state variable}
\State  \quad  $\vec a_{t+1} = (1 - \gamma) \, \vec a_{t|t} + \gamma \, \vec q$
\State \quad  $P_{t+1} = (1 - \gamma)^2 P_{t|t} + \gamma^2 Q$
\end{algorithmic}
\end{algorithm}

The initial states and hyperparameters play a crucial role in the Kalman filter. 
To determine these values, it is essential to leverage available professional knowledge and perform manual tuning. 
All the required initial states and parameters are outlined and introduced below. Further explanations, numerical results, and recommendations for selecting their values are provided in Section~\ref{ss:para_kf}.

    \begin{itemize}
        \item Initial states
            \begin{itemize}
                \item Mean vector of the initial state variable $\vec a_1$. According to equations~\eqref{eq:lim_a_tp1} and~\eqref{eq:eq_state_cuni}, the mean value of the fraction will converge to \( \vec q \), so it is reasonable to set \( \vec a_1 = \vec q \) as the initial mean value. These values can be estimated using expertise or through ordinary least squares estimation from recent historical data.
                
                \item Covariance matrix $P_1$ of the initial state variable $\vec \alpha_1$. According to equations~\eqref{eq:lim_P_tp1} and~\eqref{eq:eq_state_cuni}, the initial value $P_1$ does not matter too much, and the variance of \( \vec \alpha_t \) will converge to $P_\infty = (\gamma / (2- \gamma))\, Q$. To simplify, when applying the Kalman filter,  we first set the value of $P_\infty$, and then choose the value of $\gamma$, from which $Q$ is computed. The initial covariance matrix is then set to $P_1 = Q$.       
            \end{itemize}
            
        \item Parameters
            \begin{itemize}
                \item Process noise mean vector $\vec q$.
                The value of $\vec q$ can be estimated using expertise or through ordinary least squares estimation from recent historical data.

                \item Time step correlation parameter $\gamma$. 
                The value of $\gamma$ and $Q$ are not independent. Thus in practice, based on equation~\eqref{eq:lim_P_tp1}, we first set the value of $P_\infty$, fix one value for $\gamma$, and then compute $Q$ accordingly.
                
                \item Process noise covariance $Q$.  Similarly, this can be calculated using equation~\eqref{eq:lim_P_tp1}, given $\gamma$ and $P_\infty$. More details will also be explained in Subsection~\ref{sss:para_kf_var}.

                \item Covariance matrix of measurement noise $H_t$.
                Since the observation variable is a scalar in the model, \( H_t \) is also a scalar and represents the variance of observation noise. The value of \( H_t \) can be estimated based on prior knowledge. For example, for elements such as Cu, according to equation~\eqref{eq:estimate_Ht}, prior knowledge about \( m_{\text{steel}} \) and \( m_{\text{hm}} \) from the capacities of the basic oxygen furnace or blast furnace, as well as information on the measurement errors for \( f_{\text{steel}, X} \) and \( f_{\text{hm}, X} \), can be used to estimate the value of \( H_t \). 
            \end{itemize}
    \end{itemize}
    
\subsubsection{Non-linear state space model: unscented Kalman filter}
The model for elements like Cr and S, equations~\eqref{eq:eq_obs_crs}--\eqref{eq:eq_state_crs}, is a more complicated non-linear state space model because of the occurrence of the function $Z_t(\vec \alpha_t^+)$ in~\eqref{eq:zt_crs}. 

Fortunately several approaches have been developed to fit such models. In our work, we use the \textit{Unscented Kalman Filter (UKF)} as explained in~\cite[Section 10.3]{durbin2012time}. 
The unscented Kalman filter is an algorithm used to estimate the state of a non-linear non-Gaussian model from noisy observations.
Unlike the extended Kalman filter (EKF)~\cite{ribeiro2004kalman}, which linearizes the system using first-order Taylor expansions, the UKF avoids explicit linearization and instead applies the unscented transform~\cite{angrisani2005unscented}. The key idea behind this transform, is that it is more accurate to approximate how a distribution propagates through a non-linear function than to approximate the function itself. To achieve this, the UKF selects a set of carefully chosen sample points, known as sigma points, around the mean of the state estimate. These sigma points are propagated through the true non-linear function, rather than an approximation. By evaluating how the sigma points evolve, the UKF captures higher-order effects of non-linearity that the EKF would otherwise miss. The variance and covariance of the transformed points are then recomputed to approximate the posterior state distribution. This allows the UKF to provide more accurate state estimates without requiring explicit Jacobians or Hessians.

We need to mention that 
our model~\eqref{eq:eq_obs_crs}--\eqref{eq:eq_state_crs} differ from the model in~\cite[(9.36) and (9.37)]{durbin2012time}. It is assumed that the mean of the process noise distribution is 
$[\vec q, \, \vec q_c ]^\top$ rather than $0$, which causes small modifications in the algorithm.  

For completeness, the full algorithm is provided in Algorithm~\ref{alg:km_crs}, with sigma points generated using Algorithm~\ref{alg:sigma_pts}. 
It is worth mentioning that there are numerous other approaches to implement the UKF. Various UKF variants available in the literature are discussed and compared in~\cite{menegaz2015systematization}.

For convenience, we summarize the different initial values and parameters required for the UKF. 
\begin{itemize}
    \item Initial states
            \begin{itemize}
                \item Mean vector $\vec a_1^+$ of initial state variable $\vec \alpha_1^+$. According to equations~\eqref{eq:lim_a_tp1} and~\eqref{eq:eq_state_crs}, the mean value of the $\vec \alpha_t^+$ will converge to \([ \vec q, \, \vec q_c]^\top\).
                
                \item Covariance matrix $P_1^+$ of initial state variable $\vec \alpha_1^+$. According to equations~\eqref{eq:lim_P_tp1} and~\eqref{eq:eq_state_crs}, the initial value $P_1$ does not matter too much, and the variance of \( \vec \alpha_t \) will converge to $ P_\infty^+ = (\gamma / (2- \gamma))\, \begin{bmatrix} Q & 0 \\ 0 & Q_c \end{bmatrix}$.
            \end{itemize}
            
        \item Parameters
            \begin{itemize}
                \item Process noise mean vector $\vec q$ and $\vec q_c$.
                
                \item Process noise covariance $Q$ and $Q_c$.

                \item The length of $\vec \alpha_t^+$ is $m$, which equals the sum of the number of scrap types and the number of additional parameters used to estimate the partition coefficients, as defined in~\eqref{eq:linear_ell_t}.

                \item An additional parameter $k$  controls the spread of the sigma points, which provides an extra degree of freedom to tune the higher-order moments of the approximation and can be used to reduce the overall prediction errors~\cite{julier1997new}.
                When the state variable is assumed Gaussian, a useful heuristic is to select $m+k = 3$. The results for our case can be found in Section~\ref{sss:para-ukf-k}.
            \end{itemize}
\end{itemize}

\begin{adjustwidth}{-5em}{-5em}
\noindent\begin{minipage}{1.0\linewidth}
\begin{algorithm}[H]
\caption{UKF\_Step (($\vec m_t, \,m_{\rm steel, t},\, f_{{\rm steel}, X, t},\, m_{{\rm hm}, t},\, f_{{\rm hm}, X, t},\, m_{{\rm slag}, t},\, f_{{\rm FeOn, slag}, t}$); ($\vec a_t^+, P_t^+$);  $\gamma$, $\vec q$, $\vec q_c$, $Q$, $Q_c$, $H_t$)}
\label{alg:km_crs}
\begin{algorithmic}
\State $y_t =m_{{\rm steel}, X, t} f_{{\rm steel}, X, t}$
\State $Z_t (\vec a^+) = Z_t (\vec a, \vec c) = 
        \frac{
            \vec m_t \cdot \vec a + m_{\text{hm}, t} \, f_{\text{hm}, t}
        }{
            1 + (c_{1} + c_{2}  f_{\text{FeOn, slag}, t}) \, m_{\text{slag}, t} / m_{\text{steel}, t} 
        }$

\vspace{0.5em}
\State \textbf{Sigma point preparation}
\State \quad Compute Cholesky decomposition: $P_t^+ = A_t A_t^\top$
\State \quad $m = {\rm length}\, (\vec a_t^+)$
\State \quad $\{ (\vec x_{t},\, w_{t})_i\}_{i=-m}^m = {\rm sigma\_points}(\vec a_t^+, A_t)$

\vspace{0.5em}
\State \textbf{Compute predicted output and covariances}
\State \quad $\bar{y}_t = \sum_{i=-m}^{m} w_i Z_t(\vec x_{t,i})$
\State \quad $P_{a v, t} = \sum_{i=-m}^{m} w_i \big(\vec x_{t,i} - \vec a_t^+ \big) ( Z_t(\vec x_{t,i}) - \bar{y}_t )$
\State \quad $P_{vv, t} = \sum_{i=-m}^{m} w_i ( Z_t(\vec x_{t,i}) - \bar{y}_t )^2 + H_t$
\State \quad $v_t = y_t - \bar{y}_t$

\vspace{0.5em}
\State \textbf{Update state variables based on new measurement $(\vec m_t, y_t)$}
\State \quad $\vec a_{t|t}^+ = \vec a_t^+ +  P_{a v, t} P_{vv,t}^{-1} v_t$
\State \quad $P_{t|t}^+ = P_t^+ - P_{a v, t} P_{vv,t}^{-1} P_{a v,t}^\top$

\vspace{0.5em}
\State \textbf{Predict the next state variables}
\State \quad $\vec a_{t+1}^+ = (1 - \gamma) \, \vec a_{t|t}^+ + \gamma \, [ \vec q,\,\vec q_c]^\top$
\State \quad $P_{t+1}^+ = (1 - \gamma)^2 P_{t|t}^+ + \gamma^2 \begin{bmatrix} Q & 0 \\ 0 & Q_c\end{bmatrix}$
\end{algorithmic}
\end{algorithm}
\end{minipage}
\end{adjustwidth}

\begin{algorithm}[h!]
\caption{sigma\_points($\vec a$, $A$) }
\label{alg:sigma_pts}
\begin{algorithmic}
\State $k=3$
\State $m = {\rm length}(\vec a)$
\State $\vec x_{0} = \vec a,  \quad w_0 = k/ (m + k)$
\State $\vec x_{\pm i} = \vec a \pm \sqrt{m + k A_{:,i}},   \quad w_{\pm i} = 1 / (2(m + k)), \quad {\rm for}\quad  i=1,\ldots,m$
\State return $\{ ( \vec x, \,w)_i \}_{i=-m}^m$
\end{algorithmic}
\end{algorithm}

\subsection{Hyperparameters in the Kalman filter}
\label{ss:para_kf}
\begin{figure}[htbp]
\centering
\begin{adjustwidth}{-4em}{-4em} 
\centering
\subfloat[\centering With true hyperparameters.]{\includegraphics[width=0.4\linewidth]{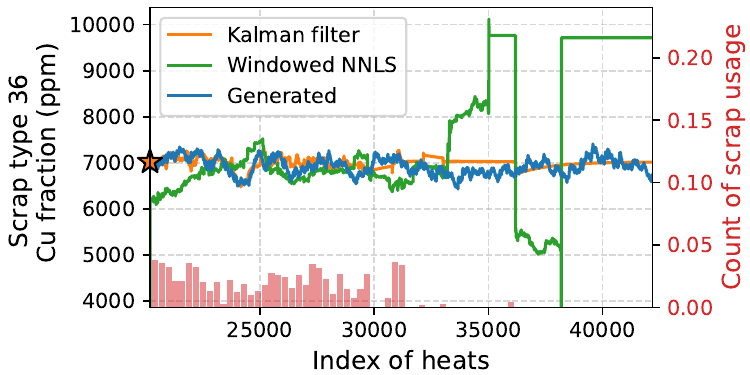} \label{fig:cu_true_f}}
\subfloat[\centering With true hyperparameters.]{\includegraphics[width=0.4\linewidth]{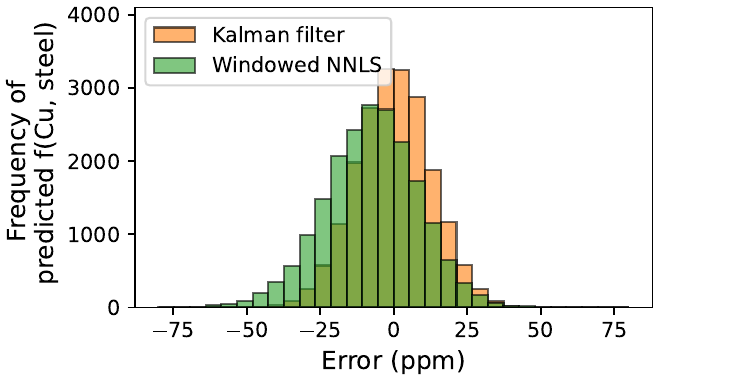} \label{fig:cu_true_output}}\\
\centering
\subfloat[\centering With $\vec q = 0.5 \,\vec a_1^{\,\rm true}$.]{\includegraphics[width=0.4\linewidth]{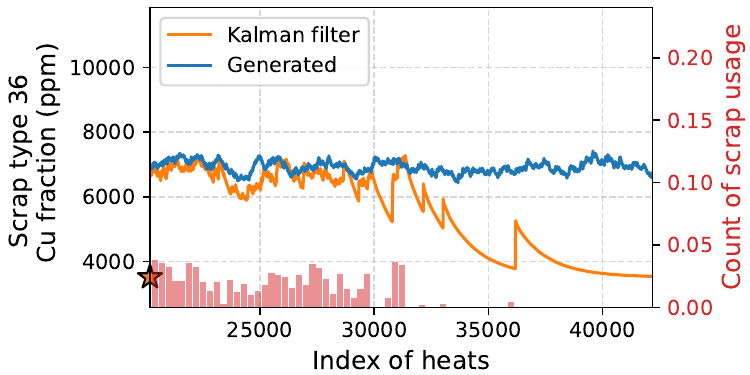} \label{fig:cu_small_mean_f}}
\subfloat[\centering With $\vec q = 1.5 \,\vec a_1^{\,\rm true}$.]{\includegraphics[width=0.4\linewidth]{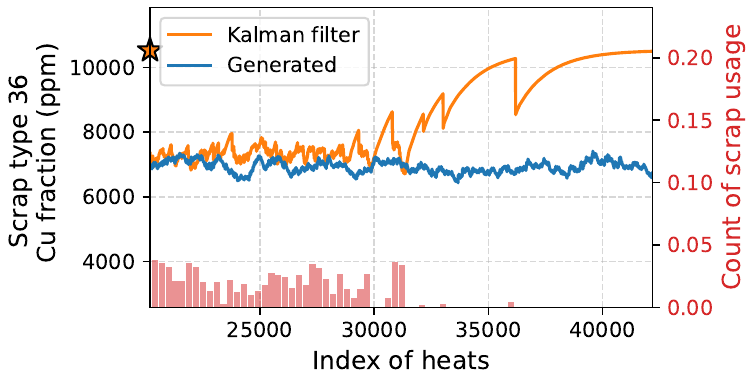} \label{fig:cu_large_mean_f}}\\
\centering
\subfloat[\centering With $H_t = 0.01 \,H^{\,\rm true} $.]{\includegraphics[width=0.4\linewidth]{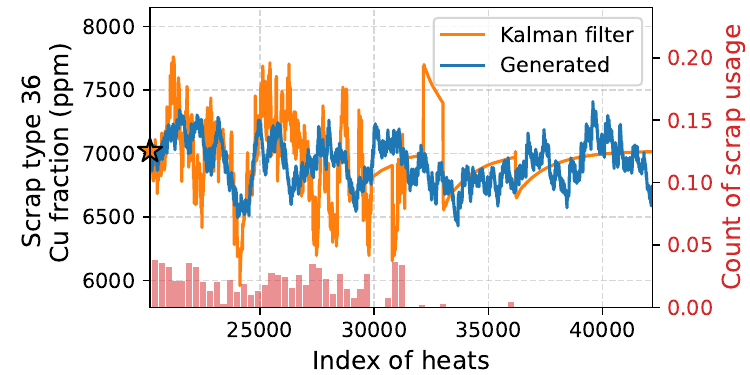} \label{fig:cu_small_obs_f}}
\subfloat[\centering With $H_t = 25.0 \,H^{\,\rm true} $.]{\includegraphics[width=0.4\linewidth]{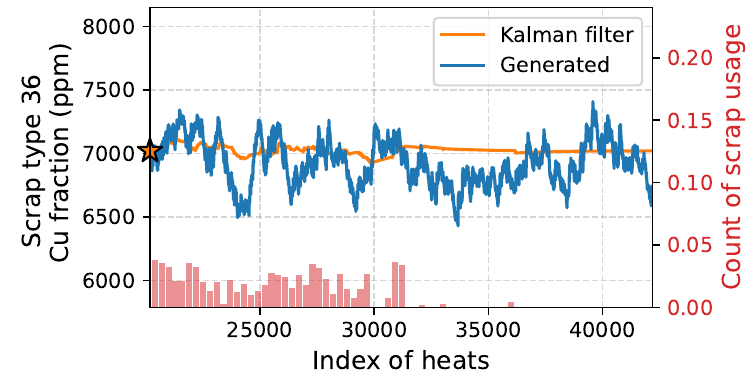} \label{fig:cu_large_obs_f}}\\
\centering
\subfloat[\centering With $\gamma = 0.01\, \gamma^{\,\rm true}$ and $Q = Q^{\,\rm true}$.]{\includegraphics[width=0.4\linewidth]{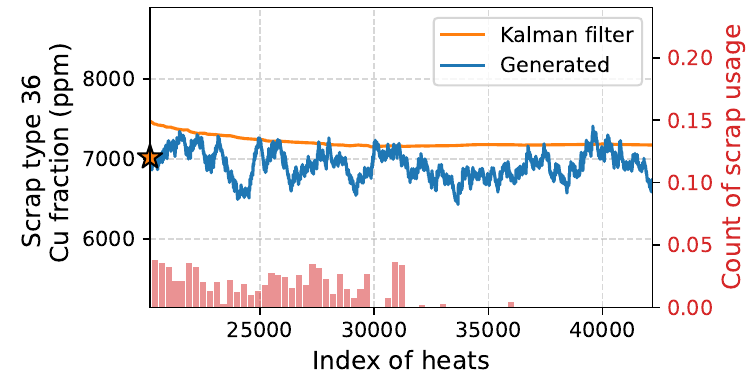} \label{fig:cu_small_gamma_f}}
\subfloat[\centering With $\gamma = 25.0\, \gamma^{\,\rm true}$ and $Q = Q^{\,\rm true}$.]{\includegraphics[width=0.4\linewidth]{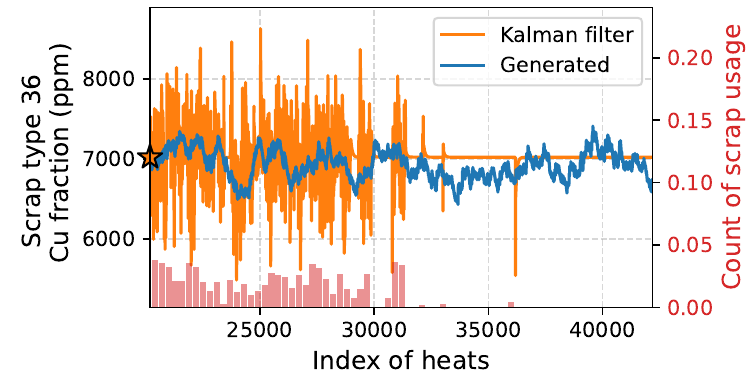} \label{fig:cu_large_gamma_f}}\\
\centering
\subfloat[\centering With $\gamma = 0.01 \, \gamma^{\,\rm true}$ and $P_\infty = P_\infty^{\rm true}$.]{\includegraphics[width=0.4\linewidth]{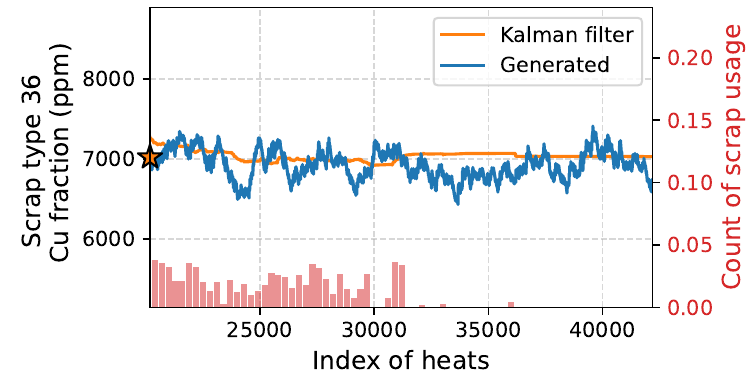} \label{fig:cu_small_gamma_Q_f}}
\subfloat[\centering With $\gamma = 25.0 \, \gamma^{\,\rm true}$ and $P_\infty = P_\infty^{\rm true}$.]{\includegraphics[width=0.4\linewidth]{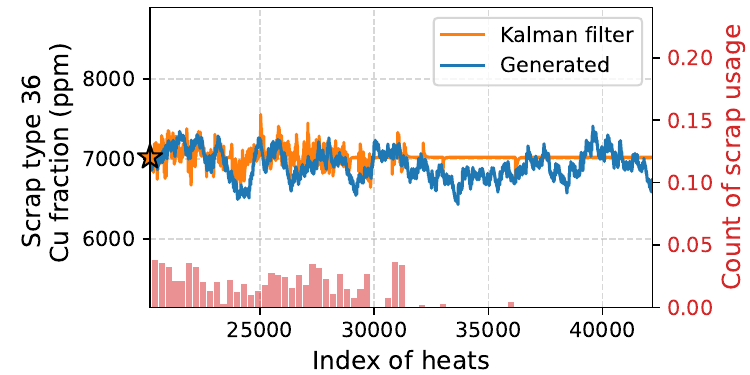} \label{fig:cu_large_gamma_Q_f}}
\caption{
Effects of misspecified hyperparameters in the Kalman filter for scrap type $36$, which is rarely used in the second half of the heats. 
Figures~\ref{fig:cu_true_f}--\ref{fig:cu_true_output} use the true hyperparameters. 
Figures~\ref{fig:cu_small_mean_f}--\ref{fig:cu_large_mean_f} show the impact of misspecification in the mean vector \( \vec{q} \), causing shifts in estimated Cu when no observations are available. 
Figures~\ref{fig:cu_small_obs_f}--\ref{fig:cu_large_obs_f} illustrate the effect of misestimating the observation covariance \( H_t \), which governs trust in the data. 
Figures~\ref{fig:cu_small_gamma_f}--\ref{fig:cu_large_gamma_Q_f} address misspecification in \( \gamma \), \( Q \), and \( P_\infty \), which together control the rate and variability of the estimated Cu fraction. 
}
\label{fig:cu_group_results}
\end{adjustwidth}
\end{figure}

{We have introduced the Kalman filter and UKF for linear and non-linear models, respectively, along with the required initial states and hyperparameters.  We now return to a more detailed discussion of the Kalman filter. The corresponding details for the UKF will be presented in Section~\ref{ss:para-ukf}. Specifically,  we are going to discuss the influence of misspecification of the parameters in the linear model~\eqref{eq:eq_obs_cuni}--\eqref{eq:eq_state_cuni}, and give some practical suggestions about how to choose the parameters. To make it clear, the key parameters are listed below:
\begin{itemize}
    \item \( \vec q \), \( \vec a_1 \): The mean value of the fraction, with \( \vec a_1 = \vec q \).
    \item \( H_t \): The observation noise covariance matrix.
    \item \( \gamma \): The correlation parameter between fractions at consecutive time steps.
    \item \( Q \), \( P_1 \): The covariance matrix $P_1$ of the initial state $\vec \alpha_1$ is always set the same as the process noise covariance matrix $Q$.
\end{itemize}
While plausible values for these parameters were used during data generation, they must be estimated in practice. We will demonstrate how these parameters influence the results of the Kalman filter when set correctly or incorrectly.

This subsection uses Cu as an example, with generated data as the true values and windowed non-negative least squares (NNLS) regression~\cite{slawski2013non} for comparison. Windowed NNLS estimates non-negative parameters using a sliding window of data, adapting to time-varying conditions, which is suitable for dynamic systems like steel production. Although ordinary least squares (OLS) can estimate the Cu fraction, the time-varying and non-negative nature of the steel process requires the use of windowed NNLS.
To determine the optimal window size, some numerical experiments were conducted. Rather than concentrating on the Cu fraction of a specific scrap type, we direct our attention to the error in predicting the  Cu fraction of steel. This is also of practical use as in reality this is the only thing which can be measured. The results indicate that smaller window size reduce bias but increase the standard deviation of the error. As the window size grows, both the mean and standard deviation of the error tend to stabilize. When the window size exceeds $1700$, its impact on the bias and variance of the error becomes minimal.
While the window size can be adjusted, increasing it results in the estimated Cu fraction converging towards the mean value of using all of the data and slowing its rate of change, presenting a trade-off between stability and responsiveness.
Thus the window size is set to a reasonable value of $2000$ heats.

When the hyperparameters match the data generation model, the results are shown in Figure~\ref{fig:cu_true_f}--\ref{fig:cu_true_output}. 
Figure~\ref{fig:cu_true_f} illustrates the Cu fraction for scrap type 36. The orange line represents the Kalman filter results as proposed in the paper, the green line shows the windowed NNLS results, and the blue line indicates the true generated values. The orange star marks the corresponding component in the mean vector $\vec q$ for the Kalman filter. At the bottom of the figure, red bars indicate the usage frequency of scrap types. We remind the readers that this part of the generated data comes from actual production data. Scrap type 36 is frequently used in the first half of the heats but rarely in the latter half. As expected, when}this scrap type is unused, the windowed NNLS results fluctuate significantly and rapidly, while the Kalman filter remains stable near the mean vector $\vec q$, demonstrating more plausible performance.

Figure~\ref{fig:cu_true_output} displays the error between the measured Cu fraction $f_{{\rm steel, Cu}, t}$ and the estimated Cu fraction $\hat{f}_{{\rm steel, Cu}, t}$ in steel at $t = T_0,\ldots, T$, where $T_0 = 5001$  and $T = 20000$ denotes the total number of heats.  
The first 5000 heats are excluded from the error calculation, as they are used to estimate the initial state of the Kalman filter. Similarly, the windowed NNLS method requires the first $2000$ heats to perform its first estimation. For consistency, all the following experiments in this paper also exclude the first $5000$ heats. The Kalman filter yields a mean error of  $-0.06$ ppm with a standard deviation of $13.25$ ppm, while the windowed NNLS results in a mean error of $-7.56$ ppm and a standard deviation of $16.93$ ppm. Although the errors are relatively similar, the Kalman filter provides significantly better estimates of the scrap composition.
Other numerical results are also shown in Figure~\ref{fig:cu_group_results} and will be discussed in detail in the following subsections.

\subsubsection{Mean}
\label{sss:para_kf_mean}
To explore the impact of the mean vector \( \vec q \) and  the initial state \( \vec a_1 \), we fix all other parameters as in the data generation process and vary \( \vec q \) and \( \vec a_1 \) from \( 0.5 \, \vec a_1^{\,\rm true} \) to \( 1.5 \, \vec a_1^{\,\rm true} \). Some of the  true values of the mean vector $\vec a_1^{\,\rm true}$ are shown in Table~\ref{tab:mean_concentration}.
After estimating the state variable, which is the element fraction of scrap, the fraction of elements in steel can be calculated and compared with the measured values. 

Table~\ref{tab:para_cu_mean} shows the statistical properties of the error in predicting the Cu fraction of steel
\begin{equation}
    e_t = \hat{f}_{{\rm steel, Cu}, t} - f_{{\rm steel, Cu}, t}, \quad t = T_0, \ldots, T,
    \label{eq:err_f_steel_cu}
\end{equation} 
The mean and standard deviation of the errors $e_t$ are calculated using

\begin{equation}
    \bar{e} = \frac{1}{T - T_0 + 1} \sum_{t=T_0}^{T} e_t, \quad   
    {\mathrm{Var}}[e] = \frac{1}{T - T_0} \sum_{t=T_0}^{T} (e_t - \bar{e})^2.
    \label{eq:err_f_steel_cu_meanvar}
\end{equation}

Figure~\ref{fig:cu_small_mean_f}--\ref{fig:cu_large_mean_f} show the influence of misspecified mean vector for scrap type $36$ which is seldom used in the second half of the heats. In Figure~\ref{fig:cu_small_mean_f}, the mean vector is set to \( \vec q = 0.5 \, \vec a_1^{\,\rm true} \), while in Figure~\ref{fig:cu_large_mean_f}, \( \vec q = 1.5 \, \vec a_1^{\,\rm true} \).
The orange lines show the estimated Cu fraction using the Kalman filter, while the blue lines correspond to the generated values.
The results in the table and the figures indicate that as the mean vector deviates from the true value, the mean error shifts in the same direction. This can be explained by the fact that when a scrap type is unused, the estimated fraction will approach the mean vector \( \vec q \).

In conclusion, when estimating impurity fractions, it is advisable to set a slightly larger mean vector to ensure the products remain safe, avoiding impurity levels above the threshold. However, the mean vector should not be excessively large and should be based on expertise or through  estimation from historical data.

\begin{table}[ht]
\centering
\caption{Kalman filter results for the error of predicted Cu fraction in steel with different mean vectors (unit: ppm), based on equation~\eqref{eq:err_f_steel_cu} and~\eqref{eq:err_f_steel_cu_meanvar}.}
\label{tab:para_cu_mean}
\begin{tabular}{SSS} \toprule
{mean vector} & {mean(error) }   & {std(error)  }  \\  \midrule 
{$0.5 \, \vec a_1^{\,\rm true}$}  & {$-9.47$} & {16.85} \\ 
{$0.7 \, \vec a_1^{\,\rm true}$}  & {$-5.58$}  & {14.09}  \\ 
{$1.0 \, \vec a_1^{\,\rm true}$}  & {$-0.06$}  & {13.25}  \\ 
{$1.2 \, \vec a_1^{\,\rm true}$}  & {$ 3.59$}  & {13.57}  \\ 
{$1.5 \, \vec a_1^{\,\rm true}$}  & {$ 9.34$}  & {16.87}  \\ \bottomrule
\end{tabular}
\end{table}

\subsubsection{Observation noise }
To assess the impact of \( H_t \), which is kept constant, we keep all parameters the same as the generated true values, except for \( H_t \). The true value of \( H_t \) is denoted by \( H^{\,\rm true} \). We vary \( H_t \) from \( 0.01\, H^{\,\rm true} \) to \( 25.0\, H^{\,\rm true} \). 

Table~\ref{tab:para_cu_H} shows the Kalman filter results with different values of \( H_t \). The last column of the table shows how fast the estimated Cu fraction of scrap type 36 changes.
The estimated Cu fraction of scrap type 36 is shown in Figure~\ref{fig:cu_small_obs_f}--\ref{fig:cu_large_obs_f}.
In Figure~\ref{fig:cu_small_obs_f}, the observation covariance matrix is set to \( H_t = 0.01 \, H^{\,\rm true} \), while in Figure~\ref{fig:cu_large_obs_f}, it is set to \( H_t = 25.0 \, H^{\,\rm true} \). 

The results show that the value of \( H_t \) has a minimal effect on the predicted element fraction in steel but significantly influences the estimation of the fraction in scrap. When \( H_t \) is smaller than the true value, the Kalman filter over-trusts the observations, causing the estimated state variable to change more rapidly than the true value. However, when $H_t$ is larger, the filter under-trusts the observations, leading to a state variable that tends to center around the mean value from the left panel.
Since observation noise has a physical meaning, it is best to estimate \( H_t \) based on physical knowledge and expert judgment, such as in equation~\eqref{eq:estimate_Ht}.

\begin{table}[htbp]
\centering
\caption{Kalman filter results for the predicted Cu fraction in steel with different observation covariance matrices  (unit: ppm), based on equation~\eqref{eq:err_f_steel_cu} and~\eqref{eq:err_f_steel_cu_meanvar}. Also shown is the standard deviation of the predicted Cu fraction for scrap type $36$, averaged over time. This scrap type is scarcely used in the second half of the heats.}
\label{tab:para_cu_H}
\begin{tabular}{SSSS}
    \toprule
    {$H_t$} & {mean(error)}    & {std(error)}  & {std($f_{\rm Cu, 36}$)}   \\ \midrule
    {$0.01\, H^{\,\rm true}$ }    & {$-0.43$}   & {14.47}  & {274} \\
    {$0.1\, H^{\,\rm true}$  }    & {$-0.05$}   & {13.42}  &{193} \\
    {$1.0 \, H^{\,\rm true}$ }    & {$-0.06$}  & {13.25}  & {113}\\
    {$9.0 \, H^{\,\rm true}$ }    & {$-0.13$}   & {13.47}  & {48}\\
    {$25.0 \, H^{\,\rm true}$}   &  {$-0.24$}  &  {13.69}  & {31} \\ \bottomrule
    \end{tabular}
\end{table}

\subsubsection{Variance}
\label{sss:para_kf_var}
According to equation~\eqref{eq:lim_P_tp1}, \( P_\infty \), \( \gamma \), and \( Q \) are interrelated. The final uncertainty \( P_\infty \) is determined by \( \gamma \) and \( Q \). Thus, we will:
\begin{itemize}
    \item Fix \( Q \) and observe how \( P_\infty \) and the results change with \( \gamma \).
    \item Fix \( P_\infty \) and observe how the results change with \( \gamma \).
\end{itemize}

\paragraph{\bf{Fix \( Q \) and vary \( \gamma \).}}  
The true value for data generation is \( \gamma^{\,\rm true} = \ln 2 / 1000 \), and in the experiments, we vary \( \gamma \) from \( 0.01 \, \gamma^{\,\rm true} \) to \( 25.0 \, \gamma^{\,\rm true} \). The predicted element fraction in steel is compared with the measured values, with the mean and standard deviation of the error shown in Table~\ref{tab:para_cu_gamma}.
As seen in the table, the mean and standard deviation of the error remain relatively stable across different values of \( \gamma \). However, varying \( \gamma \) changes \( P_\infty \), which influences the rate and range at which the state variable changes. Figure~\ref{fig:cu_small_gamma_f}--\ref{fig:cu_large_gamma_f} show how the estimated state variable changes with different \( \gamma \) values  and $Q = Q^{\rm true}$.
Figure~\ref{fig:cu_small_gamma_f} shows the results for \( \gamma = 0.01 \, \gamma^{\,\rm true} \), where the state variable changes too slowly. In contrast, Figure~\ref{fig:cu_large_gamma_f} shows results for \( \gamma = 25.0 \, \gamma^{\,\rm true} \), where the state variable changes too rapidly between a large range, from $5000$ and $8000$ ppm, which is unrealistic.
Thus, while \( \gamma \) does not significantly affect the element fraction in steel, it impacts the rate and range at which the estimated state variable evolves. There are some spikes in the second half of the heats in Figure~\ref{fig:cu_large_gamma_f}, because the scrap type $36$ is used near the spikes.

\begin{table}[htbp]
\centering
\caption{Kalman filter results for the error of predicted Cu fraction in steel  with different \( \gamma \) values (fix \( Q = Q^{\,\rm true} \)) (unit: ppm).}
\label{tab:para_cu_gamma}
\begin{tabular}{SSSS}
    \toprule
    {$\gamma$}           & {$P_\infty$}  & {mean}     & {std}     \\ \midrule
    {$0.01\, \gamma^{\,\rm true}$}  & {$0.01 P_\infty^{\,\rm true}$}  &  {$0.64$}  &  {14.24} \\ 
    {$0.1 \, \gamma^{\,\rm true}$}  & {$0.10 P_\infty^{\,\rm true}$ }  &  {$-0.06$}  &  {13.77} \\ 
    {$1.0 \, \gamma^{\,\rm true}$}  & {$P_\infty^{\,\rm true}$         }  &  {$-0.06$}  &  {13.25} \\ 
    {$9.0 \, \gamma^{\,\rm true}$}  & {$9.03 P_\infty^{\,\rm true}$}  &  {$-0.13$}  &  {13.73} \\ 
    {$25.0\, \gamma^{\,\rm true}$}  & {$25.21 P_\infty^{\,\rm true}$}  &  {$-0.16$}  &  {14.57} \\ 
    \bottomrule
\end{tabular}
\end{table}

\paragraph{{\bf Fix \( P_\infty \) and vary \( \gamma \).}  }
According to equation~\eqref{eq:lim_P_tp1}, if we fix \( P_\infty = P_\infty^{\,\rm true} \) and vary \( \gamma \) from \( 0.0001 \, \gamma^{\,\rm true} \) to \( 25.0 \, \gamma^{\,\rm true} \), we can calculate \( Q = ((2 - \gamma) / \gamma) \, P_\infty^{\,\rm true} \).
The results, shown in Table~\ref{tab:para_cu_gamma_Q}, indicate that the mean and standard deviation of the error remain relatively constant across different values of \( \gamma \).

Figure~\ref{fig:cu_small_gamma_Q_f}--\ref{fig:cu_large_gamma_Q_f} depict the estimated state variable for different \( \gamma \) values (with \( P_\infty = P_\infty^{\,\rm true} \)). 
Figure~\ref{fig:cu_small_gamma_Q_f} shows the estimated Cu fraction of scrap type 36 for \( \gamma = 0.01 \, \gamma^{\,\rm true} \). Here, the estimated state variable changes slowly. In contrast, Figure~\ref{fig:cu_large_gamma_Q_f}, with \( \gamma = 25.0 \, \gamma^{\,\rm true} \), displays rapid but bounded changes in the state variable.
The range of possible changes in the estimated state variables is constrained by $P_\infty$. Although \( \gamma \) does not significantly impact the element fraction in steel, it determines the  rate at which the estimated state variable changes, while $P_\infty$ decides the allowable range of these changes.

\begin{table}[htbp]
\centering
\caption{Kalman filter results for the error of predicted Cu fraction in steel  with different \( \gamma \) values (fix \( P_\infty = P_\infty^{\,\rm true} \)) (unit: ppm).}
\label{tab:para_cu_gamma_Q}
    \begin{tabular}{SSS}
        \toprule
        {$\gamma$}           & {mean}     & {std}     \\ \midrule
        {$0.0001\, \gamma^{\,\rm true}$}   & {$0.68$} & {$14.25$} \\ 
        {$0.01  \, \gamma^{\,\rm true}$}   & {$-0.04$} & {$13.80$} \\ 
        {$1.0   \, \gamma^{\,\rm true}$}   & {$-0.06$} & {$13.25$} \\ 
        {$9.0   \, \gamma^{\,\rm true}$}   & {$-0.23$} & {$13.35$} \\ 
        {$25.0  \, \gamma^{\,\rm true}$}   & {$-0.78$} & {$13.47$} \\ 
        \bottomrule
    \end{tabular}
\end{table}

\paragraph{{\bf Brief conclusions about the influence of variance.}}
\begin{itemize}
    \item If \( Q \) is set to \( Q^{\,\rm true} \), and  \( P_\infty \) changes with \( \gamma \), the state variable changes faster as \( \gamma \) increases and spans a larger range.
    \item If \( P_\infty \) is set to \( P_\infty^{\,\rm true} \) and \( Q \) is calculated for each \( \gamma \), the state variable still changes faster as \( \gamma \) increases, but within a range determined by \( P_\infty \).
\end{itemize}
These observations of course also follow from the theory but we illustrated them here within the practicalities of the application to get some useful insights.

\subsubsection{Conclusion about parameters}
The parameters and initial values can be estimated with the following steps:
\begin{itemize}
    \item Estimate the mean vector \( \vec q \) using expertise or historical data, and set the initial mean vector \( \vec a_1 = \vec q \). For a safer estimation of impurities, set the mean vector slightly higher.
    \item Estimate the observation covariance matrix \( H_t \) using available knowledge, based on equation~\eqref{eq:estimate_Ht}.
    \item Set \( P_\infty \) to an acceptable level of variance for the state variable, based on expertise. The value of $P_\infty$ should also be based on the value of $\vec q$. Then tune \( \gamma \) manually: increase \( \gamma \) if the state variable changes too slowly, or decrease \( \gamma \) if it changes too quickly according to expert knowledge. Calculate the corresponding \( Q \) using equation~\eqref{eq:lim_P_tp1}.
\end{itemize}

\subsection{Hyperparameters in the unscented Kalman filter}
\label{ss:para-ukf}
The Kalman filter and UKF were introduced in Section~\ref{ss:alg-kf-ukf} and the hyperparameters about the Kalman filter were discussed in Section~\ref{ss:para_kf}. We now turn our attention to the hyperparameters specific to the UKF.  Since some insights from the Kalman filter analysis carry over, in this section, we focus on the impact of other parameter misspecification within the UKF and discuss how to leverage previously established results from the Kalman filter.

For the UKF, we define the following additional hyperparameters:
\begin{itemize}
    \item Mean vector of the additional state variables: $\vec{q}_c$.
    \item Variance matrix of the additional state variables: $Q_c$.
    \item Initial state estimate and covariance matrix: $\vec{a}_1^+$ and $P_1^+$.
    \item A hyperparameter for sigma points in the UKF: $k$.
    \item Observation noise covariance matrix specific for this case: $H_t$.
\end{itemize}

\begin{figure}[htbp]
\centering
\begin{adjustwidth}{-20em}{-6em} 
\hfill
\subfloat[\centering With true hyperparameters.]{\includegraphics[width=0.24\linewidth]{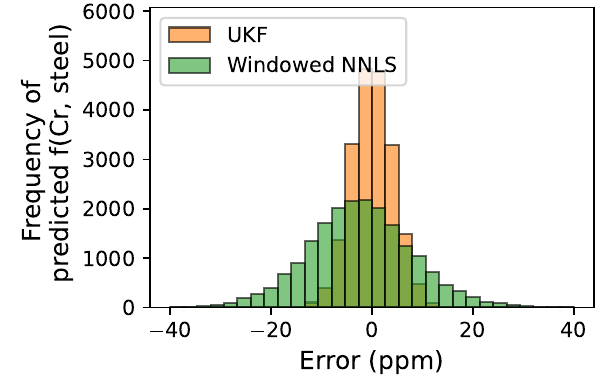} \label{fig:cr_true_output}} 
\subfloat[\centering With true hyperparameters.]{\includegraphics[width=0.3\linewidth]{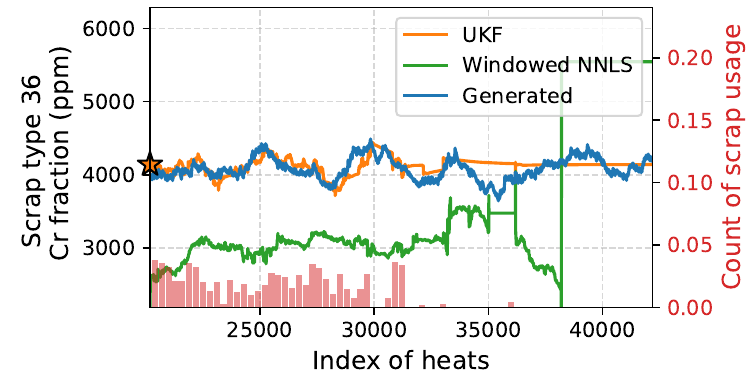} \label{fig:cr_true_value_f}}
\subfloat[\centering With true hyperparameters.]{\includegraphics[width=0.21\linewidth]{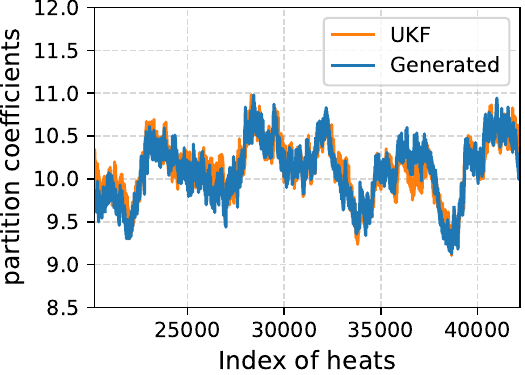} \label{fig:cr_true_value_l}}\\
\hfill
\subfloat[\centering With $\vec q_c = 0.5 \,\vec q_c^{\,\rm true}$.]{\includegraphics[width=0.3\linewidth]{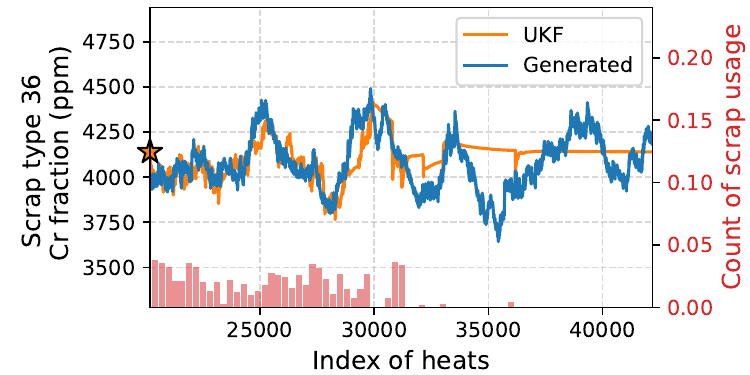} \label{fig:cr_small_mean_f}}
\subfloat[\centering With $\vec q_c = 0.5 \,\vec q_c^{\,\rm true}$.]{\includegraphics[width=0.21\linewidth]{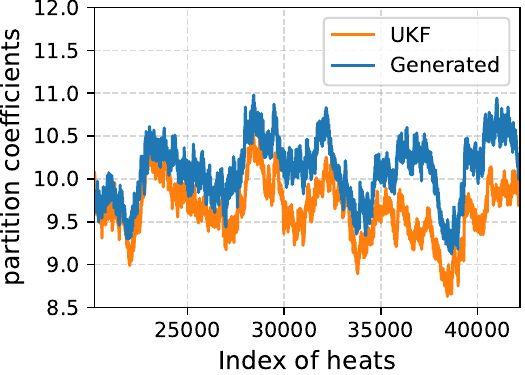} \label{fig:cr_small_mean_l}} \\
\hfill
\subfloat[\centering With $\vec q_c = 1.5 \,\vec q_c^{\,\rm true}$.]{\includegraphics[width=0.3\linewidth]{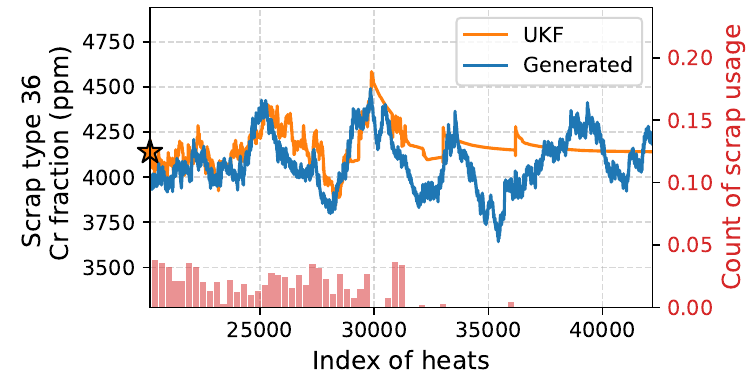} \label{fig:cr_large_mean_f}}
\subfloat[\centering With $\vec q_c = 1.5 \,\vec q_c^{\,\rm true}$.]{\includegraphics[width=0.21\linewidth]{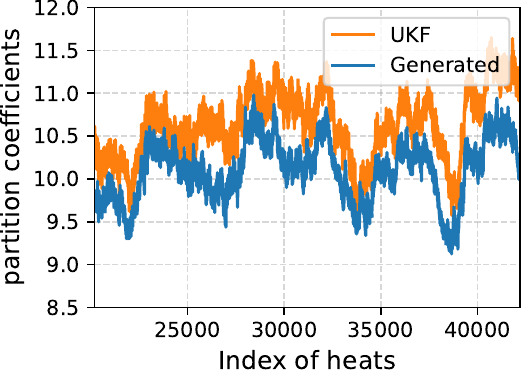} \label{fig:cr_large_mean_l}}\\
\hfill
\subfloat[\centering With  $Q_c = 0.01 \,Q_c^{\,\rm true}$.]{\includegraphics[width=0.3\linewidth]{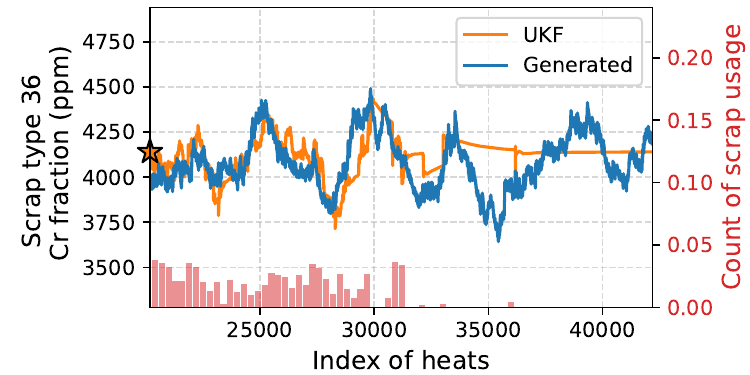} \label{fig:cr_small_cov_f}}
\subfloat[\centering With  $Q_c = 0.01 \,Q_c^{\,\rm true}$.]{\includegraphics[width=0.21\linewidth]{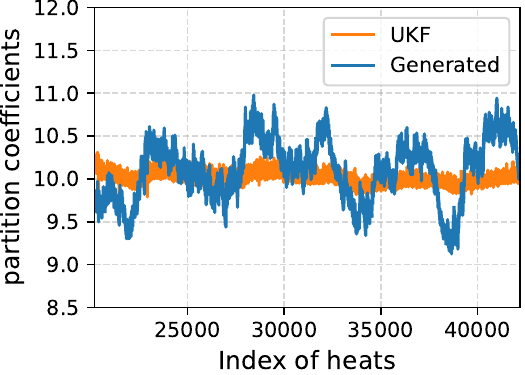} \label{fig:cr_small_cov_l}}\\
\hfill
\subfloat[\centering With $Q_c = 100.0 \,Q_c^{\,\rm true}$.]{\includegraphics[width=0.3\linewidth]{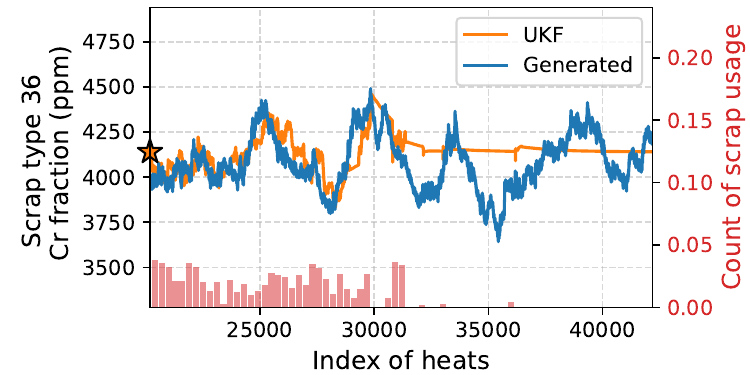} \label{fig:cr_large_cov_f}}
\subfloat[\centering With $Q_c = 100.0 \,Q_c^{\,\rm true}$.]{\includegraphics[width=0.21\linewidth]{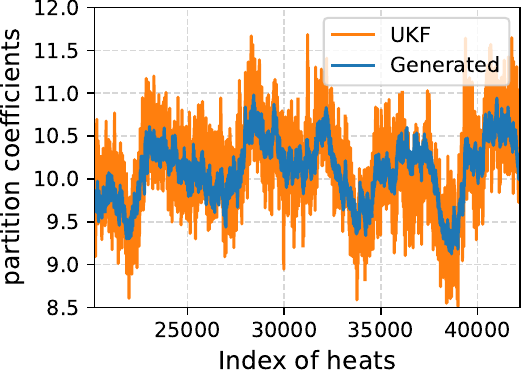} \label{fig:cr_large_cov_l}}
\end{adjustwidth}
\begin{adjustwidth}{-6em}{-6em} 
\caption{
Effects of misspecified hyperparameters for UKF. 
Figure~\ref{fig:cr_true_output}--\ref{fig:cr_true_value_l} are with the same hyperparameters when generating data. 
Figure~\ref{fig:cr_small_mean_f}--\ref{fig:cr_large_mean_l} are about misspecification on the mean vector $\vec q_c$.  Overestimating or underestimating  $\vec q_c$ leads to corresponding overestimation or underestimation of the Cr fraction and partition coefficients in the UKF results.
Figure~\ref{fig:cr_small_cov_f}--\ref{fig:cr_large_cov_l} are about misspecification on $Q_c$. Misspecification of $Q_c$ has minimal impact on the estimation of scrap composition but significantly affects the rate and range at which partition coefficients change.
The middle column shows the estimated Cr fraction of scrap type $36$, which is seldom used in the second half of heats. The right column shows the estimated partition coefficients.
}
\end{adjustwidth}
\label{fig:cr_group_results}
\end{figure}

Similarly, using equation~\eqref{eq:mass_balance_crs_f}, the predicted fraction of Cr in steel will also be used to measure the influence of misspecification. 
Since Cr distributes into both steel and slag, the problem becomes more complex. We first examine the condition without misspecification, where the UKF uses the same hyperparameters as those for data generation. The results are presented in Figure~\ref{fig:cr_true_output}--\ref{fig:cr_true_value_l}.
For the windowed NNLS, the partition coefficient is assumed constant, with $\ell_{\rm Cr}=10$. Under this assumption, equation~\eqref{eq:mass_balance_crs_f} simplifies to a linear problem. Specifically,
\[
    \sum_{s=1}^{N_{\rm s}} m_s \alpha_{s, {\rm Cr}} = f_{\rm steel, Cr} (m_{\rm steel} + m_{\rm slag} \ell_X) - m_{\rm hm} f_{\rm hm, Cr},
\]
where the right-hand side can be directly computed from available measurements. The left-hand side becomes a linear combination of the unknown scrap compositions $\alpha_{{\rm s, Cr}}$, weighted by the known input masses $m_s$. This formulation allows us to apply non-negative least squares (NNLS) within a moving time window, solving for $\alpha_{{\rm s, Cr}}$ over a limited number of heats.

Figure~\ref{fig:cr_true_output} shows the error in predicting the Cr fraction of steel. The UKF yields a mean error of $-0.02$\,ppm with a standard deviation of $4.62$\,ppm, while the windowed NNLS results in a mean error of $-2.03$\,ppm and a standard deviation fo $11.38$\,ppm. The UKF outperforms the windowed NNLS because it has both lower bias and lower standard deviation.
Figure~\ref{fig:cr_true_value_f} illustrates the Cr fraction for scrap type $36$, which is almost not used in the second half of heats, while Figure~\ref{fig:cr_true_value_l} shows the partition coefficients of Cr. 
The orange lines represent the UKF results, the green line shows the windowed NNLS results, and the blue line indicates the true generated values. The orange star marks the corresponding component in the mean vector $\vec q$ for the UKF. 
When this scrap type is not used, the windowed NNLS results fluctuate significantly and rapidly. While the UKF remains stable near the mean vector  $\vec q$, demonstrating  more reasonable performance.

Both the Cr fraction of scrap type $36$  and partition coefficients estimated by UKF follow the trend well, and the prediction for the Cr fraction in steel exhibits low bias and variance. However,  as can be seen from equation~\eqref{eq:mass_balance_crs_f}, even a small difference in the denominator can lead to significant errors in the final results. 
Table~\ref{tab:cr_true_value} shows the mean and standard deviation of the error between estimated Cr fraction in steel $\hat{f}_{\rm steel, Cr}$ using UKF and the measured Cr fraction in steel $f_{\rm steel, Cr}$, which is calculated similarly as~\eqref{eq:err_f_steel_cu} and~\eqref{eq:err_f_steel_cu_meanvar}, leaving out the first $5000$ heats.  The Cr fraction in steel is estimated using all the exact hyperparameters and hence representing the best achievable outcome.

\begin{table}[htbp]
\centering
\caption{UKF results for predicted Cr fraction in steel using equation~\eqref{eq:mass_balance_crs_f}, with the exact hyperparameters as generated data.}
\label{tab:cr_true_value}
\begin{tabular}{cccc}
\toprule
     & numerator [g] & denominator [t] & $f_{\rm Cr, steel}$ [ppm] \\ 
\midrule
mean(error) & $18$            &  $0.14$           & $-0.02$             \\ \hline
std(error)  & $1071$          &  $3.89$           & $4.62$              \\ 
\bottomrule
\end{tabular}
\end{table}

\subsubsection{Mean vector}
We have $\vec a_1^+ = [\vec q, \, \vec q_c ]^\top$, similar to Section~\ref{sss:para_kf_mean}, all other parameters are set as in the data generation process while varying $\vec q_c$  from $0.5 \, \vec q_c^{\,\rm true}$ to $1.5 \, \vec q_c^{\,\rm true}$, where $\vec q_c^{\,\rm true}$ is the vector used for generating data.  The results can be seen in Figure~\ref{fig:cr_small_mean_f}--\ref{fig:cr_large_mean_l} and Table~\ref{tab:para_cr_mean}.

It can be observed from Figure~\ref{fig:cr_small_mean_f}--\ref{fig:cr_large_mean_l} that when the mean vector for, the partition coefficients is smaller than the true value, both the fraction of Cr in the scrap types and the partition coefficients are underestimated. Conversely, when the mean vector exceeds the true value, both results are overestimated.
As shown in Table~\ref{tab:para_cr_mean}, regardless of the mean vector chosen, the error in $f_{\rm steel, Cr}$ remains largely unchanged. However, the estimation of the numerator (mass of Cr from scrap and hot metal) and the denominator (corresponding to the partition coefficients) deviates significantly from the true values. 

In summary, the model can effectively compensate for error introduced by the mean vector without notably impacting the estimation of $f_{{\rm steel, Cr}}$. This behavior can be explained by the fact that an overestimation of the Cr partition into slag results in a corresponding overestimation of the Cr mass from scrap. Therefore, it is recommended to use a larger mean vector for partition coefficients to reduce the likelihood of excessive impurities in the steel.

\begin{table}[htbp]
\centering
\caption{UKF results for the error of numerator, denominator, and predicted Cr fraction in steel with different mean vectors.}
\label{tab:para_cr_mean}

\resizebox{\textwidth}{!}{
\begin{tabular}{SSSSSSS} \toprule
{$\vec q_c$} & {$m$(num) [g]} & {$\sigma$(num) [g]} & {$m$(den) [g]} & {$\sigma$(den) [t]} & {$m$(err) [ppm]}   & {$\sigma$(err) (ppm)}  \\  \midrule 
{$0.5 \, \vec q_c^{\,\rm true}$}  & {$-1047$}         & {$1216$}         & {$-12.33$}            & {$5.70$}           & {$0.83$}           & {$4.52$}           \\
{$0.7 \, \vec q_c^{\,\rm true}$}  & {$-629$}         & {$1116$}         & {$-7.41$}            & {$4.57$}           & {$0.49$}           & {$4.61$}           \\
{$1.0 \, \vec q_c^{\,\rm true}$}  & {$10$}         & {$1049$}         & {$0.13$}            & {$3.68$}           & {$-0.03$}           & {$4.60$}           \\
{$1.2 \, \vec q_c^{\,\rm true}$}  & {$447$}         & {$1074$}         & {$5.27$}            & {$4.01$}           & {$-0.38$}           & {$4.60$}           \\
{$1.5 \, \vec q_c^{\,\rm true}$}  & {$1115$ }         & {$1212$}         & {$13.15$  }            & {$5.64$}           & {$-0.90$}           & {$4.61$}           \\ \bottomrule
\end{tabular}
}
\end{table}

\subsubsection{Variance}
Similar to Section~\ref{sss:para_kf_var}, the value of $\gamma$ is fixed to be the same as the true value, and the value of $Q_c$ varies from $0.01 \ Q_c^{\,\rm true}$ to $100 \ Q_c^{\,\rm true}$. The results can be seen in Figure~\ref{fig:cr_small_cov_f}--\ref{fig:cr_large_cov_l} and Table~\ref{tab:para_cr_cov}.

It can be observed from the figures that no matter how small or large the value of $Q_c$ is, the estimation of the fraction in scrap type aligns well with the generated values, while the estimation for the partition coefficient is significantly influenced. When $Q_c$ is too small, the estimated partition coefficients tend to be near a  constant. Conversely, when $Q_c$ is too large, the estimated partition coefficients fluctuate rapidly. The error in partition coefficients also impacts the final results for the fraction of Cr in steel.
As shown in the table, regardless of the variance matrix chosen, the error in $f_{\rm steel,\, Cr}$ remains mostly unchanged. However, the estimation of the numerator (mass of Cr from scrap and hot metal) and the denominator (related to partition coefficients) deviates a little from true values.

Thus, since the model can compensate for errors caused by the variance matrix without significantly affecting the estimation of $f_{\rm steel, Cr}$, it is reasonable to apply the same strategy as in Section~\ref{sss:para_kf_var}. Here, $P_\infty^+$ becomes:
\begin{equation}
    P_\infty^+ = \lim_{t\rightarrow \infty}{\rm Var}[\vec \alpha_{t}^+]  
    = \frac{1 - (1 - \gamma)}{1 + (1- \gamma)} 
    \begin{bmatrix}
        Q & 0 \\ 0 & Q_c
    \end{bmatrix},
    \label{eq:lim_Pplus_tp1}
\end{equation}
where $Q$ and $Q_c$ are diagonal matrices, causing $P_\infty^+$ to be a diagonal matrix, too.
Thus, $P_\infty^+$ can be initially set to an acceptable level of variance for the state variable, based on expertise or historical data. Then, $\gamma$ can be manually adjusted: increasing $\gamma$ if the state variable changes too slowly, or decreasing $\gamma$ if it changes too quickly, based on expert knowledge. The corresponding values of $Q$ and $Q_c$ can be calculated using equation~\eqref{eq:lim_Pplus_tp1}.

\begin{table}[htbp]
\centering
\caption{UKF results for the error of numerator, denominator, and predicted Cr fraction in steel with different variance matrices  (unit: ppm).}
\label{tab:para_cr_cov}
\resizebox{\textwidth}{!}{
\begin{tabular}{SSSSSSS} \toprule
{$Q_c$} & {$m$(num) [g]} & {$\sigma$(num) [g]} & {$m$(den) [g]} & {$\sigma$(den) [t]} & {$m$(err) [ppm]}   & {$\sigma$(err) [ppm]}  \\  \midrule 
    {$0.01 \, Q_c^{\,\rm true}$}  & {$-341$}          & {$1291$}         & {$-3.16$}            & {$7.98$}             & {$0.05$}           & {$4.63$}           \\
    {$0.5  \, Q_c^{\,\rm true}$}  & {$-22$}          & {$1048$}         & {$-0.22$}            & {$3.81$}             & {$-0.02$}           & {$4.60$}           \\
{$1.0 \, \vec q_c^{\,\rm true}$}  & {$10$}         & {$1049$}         & {$0.13$}            & {$3.68$}           & {$-0.03$}           & {$4.60$}           \\
    {$25.0 \, Q_c^{\,\rm true}$}  & {$44$ }          & {$1101$}         & {$0.58$ }            & {$5.91$}             & {$-0.05$}           & {$4.65$}           \\
    {$100.0\, Q_c^{\,\rm true}$}  & {$62$ }          & {$1122$}         & {$0.76$ }            & {$7.57$}             & {$-0.05$}           & {$4.75$}                           \\ \bottomrule
\end{tabular}
}
\end{table}

\subsubsection{The parameter for sigma points}
\label{sss:para-ukf-k}
The parameter $k$ controls the spread of sigma points used in the UKF, which influences how well the sigma point set captures higher-order moments of the distribution.
We set the other hyperparameters to the exact values as for generating the data and set the value of $k$ to an integer, varying from $1$ to $m-3$.  In our case,  the results are almost identical, and we can conclude that setting $k$ to an integer like $3$ is sufficient.

\subsubsection{Conclusion about parameters}
The parameters and initial values about partition coefficients can be estimated with the following steps:
\begin{itemize}
    \item Set the value of $k$ to 3.
    \item Estimate the mean vector $\vec q_c$ using expertise or historical data, and set the corresponding components in $\vec a_1^+$ the same as $\vec q_c$. For a safer estimation of impurities, set these mean vectors slightly higher.
    \item Set $P_\infty^+$ to an acceptable level of variance for the state variable, based on expertise or historical data. The value of $P_\infty^+$ should also be based on the value of $\vec q$ and $\vec q_c$. Then tune $\gamma$ manually: increase $\gamma$ if the state variable changes too slowly, or decrease $\gamma$ if it changes too quickly. Calculate the corresponding $Q$ and $Q_c$ using equation~\eqref{eq:lim_Pplus_tp1}.
    \item Similar  as in the last subsection, estimate the observation noise covariance matrix \( H_t \) based on available knowledge,
    \begin{equation*}
    {\rm Var}\left[ m_{\rm steel, \, Cr}\right]
    =
    m_{\text{steel}}^2  {\rm Var}[f_{\text{steel, Cr}}].
    \end{equation*}
\end{itemize}

\subsection{Shortcomings}
Although methods based on the Kalman filter can update the  state based on new data, there are some shortcomings we cannot ignore:
\begin{itemize}
    \item These algorithms are not totally automatic. Good estimates for the hyperparameters need to be provided, and the influence of model misspecification is shown in the previous two subsections.
    \item In the text above, we have assumed that $m_{\rm scrap}$, $m_{\rm steel}$, $m_{\rm slag}$, $m_{\rm hm}$ and $f_{\rm FeOn, \,slag}$ are measured precisely, which is not true. In reality, there will be errors for all of the measured variables. Especially, $m_{\rm slag}$, cannot be measured directly, but can only be estimated by some other measurements. The value of $m_{\rm slag}$ will influence the results of the non-linear model severely.
\end{itemize}

\section{Application on real data}
\label{s:realdata}
This section applies the Kalman filter and Unscented Kalman filter (UKF) from the previous sections to real BOF data collected from ArcelorMittal. The same approach can also be applied to EAF data by simply setting $m_{\rm hm} = 0$.

Due to the same reason as in Section~\ref{ss:para_kf}  and Section~\ref{ss:para-ukf}, the windowed non-negative least squares (NNLS) regression is used for comparison, and the window size is set to $2000$ heats. Similarly, the results about the first $5000$ heats are not shown in the figures and excluded when calculating the errors.

\begin{figure}[htbp]
\begin{adjustwidth}{-6em}{-6em} 
\centering
\subfloat[\centering ]{\includegraphics[width=7cm]{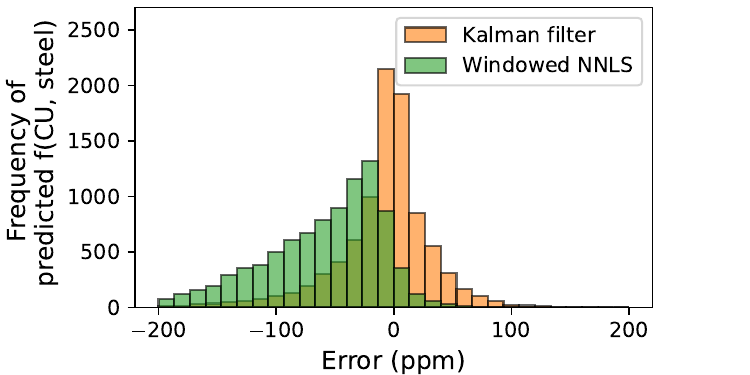} \label{fig:cu_realdata_output}}
\subfloat[\centering ]{\includegraphics[width=7cm]{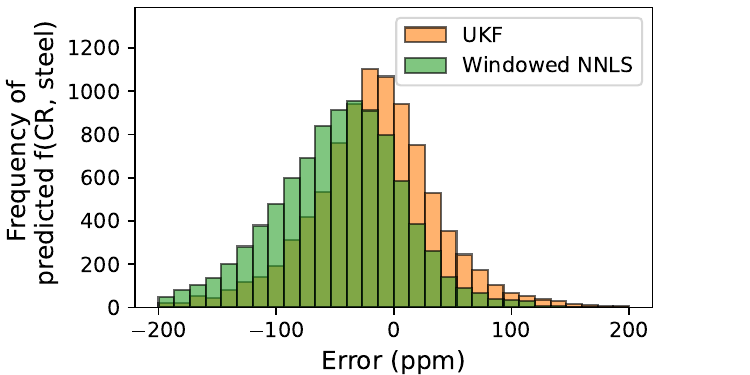} \label{fig:cr_realdata_output}}\\
\subfloat[\centering ]{\includegraphics[width=7cm]{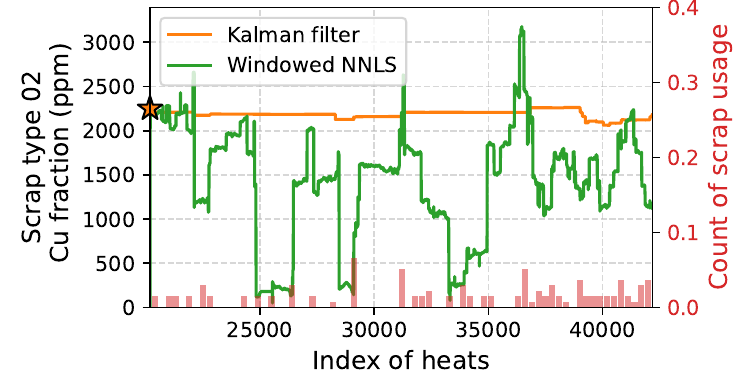} \label{fig:realdata_KF_cu2_f}}
\subfloat[\centering ]{\includegraphics[width=7cm]{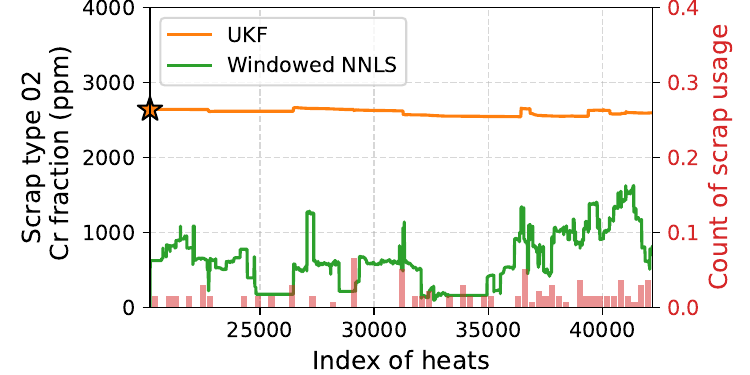} \label{fig:realdata_KF_cr2_f}}\\
\subfloat[\centering ]{\includegraphics[width=7cm]{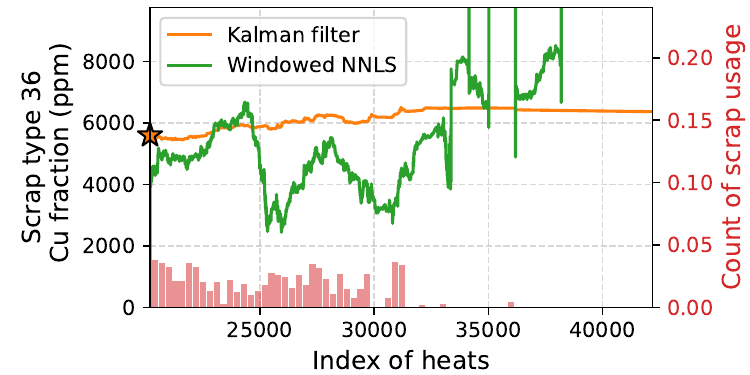} \label{fig:realdata_KF_cu36_f}}
\subfloat[\centering ]{\includegraphics[width=7cm]{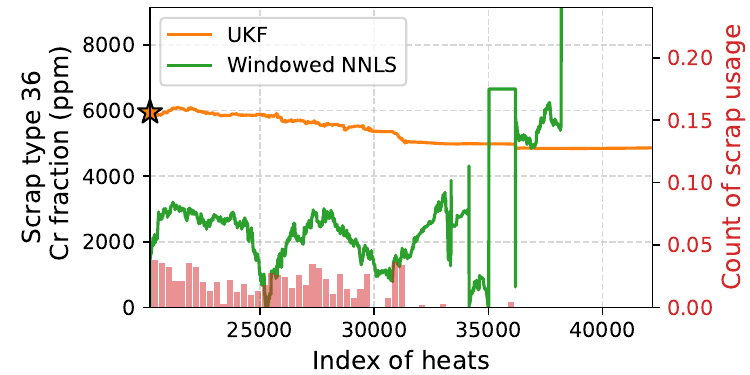} \label{fig:realdata_KF_cr36_f}}\\
\subfloat[\centering ]{\includegraphics[width=7cm]{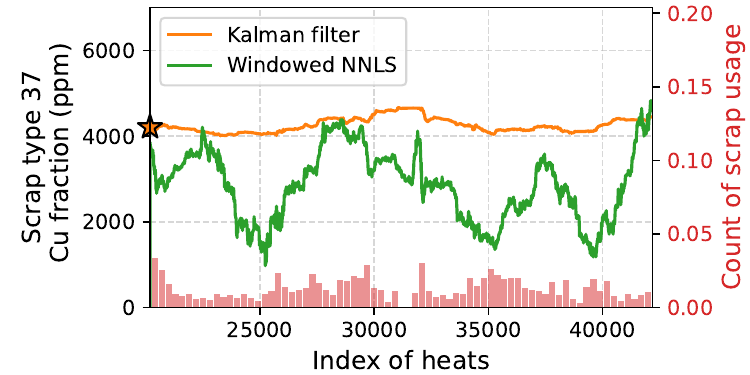} \label{fig:realdata_KF_cu37_f}}
\subfloat[\centering ]{\includegraphics[width=7cm]{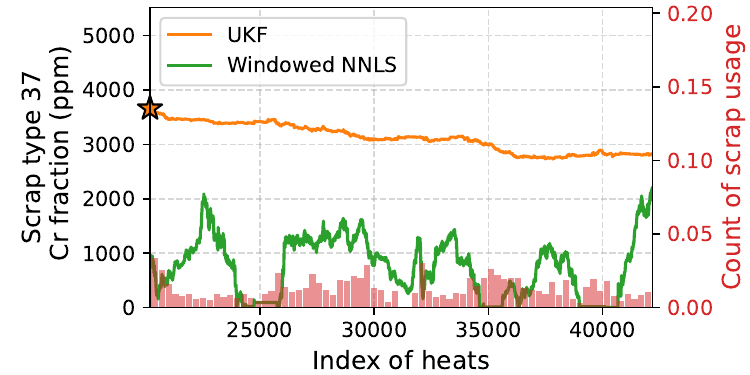} \label{fig:realdata_KF_cr37_f}}
\caption{
Results of applying windowed NNLS, Kalman filter and UKF to real data, with the left column focusing on Cu and the right column on Cr.
The first row shows the error of predicted $f_{\rm steel, Cu}$ and $f_{\rm steel, Cr}$. Both the Kalman filter and UKF outperform windowed NNLS, as they yield smaller bias and lower standard deviation of error.
The remaining figures display the estimated composition of different scrap types. The red bars at the bottom indicate scrap usage: scrap type $2$ is used infrequently, scrap type $36$ is often used in the first half of heats but rarely in the latter half, scrap type $37$ is used more frequently.
The estimated Cu fraction by the Kalman filter and the estimated Cr fraction by the UKF are more realistic compared to the results from windowed NNLS. 
}
\label{fig:data_group_results}
\end{adjustwidth}
\end{figure}

\subsection{Linear Model -- Cu}
For the Kalman filter, the mean vector $\vec{q}$ is initialized based on OLS results on the first $5000$ heats. The  matrix $P_\infty$ is diagonal with elements $(P_\infty)_{i,i} = (0.05 \, q_i)^2$. The parameter $\gamma$ is set to $7 \times 10^{-6}$, and the observation noise variance is set as $H_t = 330^2 \times 12^2 + 280^2 \times 5^2$.

It needs to be stressed that as real data lacks a ground truth of the elemental fractions in the scrap for comparison, predictions are evaluated against measured values for $f_{\rm steel, \, Cu}$. To avoid bias, only past data is used for prediction, which means no information from the heats being predicted is included, except for the input mass of scrap. 
Table~\ref{tab:ols_kalman_comparison} and Figure~\ref{fig:cu_realdata_output} display the error for predicting $f_{\rm steel, \, Cu}$. As can be seen, the Kalman filter results exhibit lower bias and standard deviation, indicating improved accuracy, compared to windowed NNLS. Compared to Figure~\ref{fig:cu_true_output}, the error distribution of the Kalman filter on real data is more peaked. This is because the generated data exhibit strong linearity, allowing windowed NNLS to predict the Cu fraction of steel reasonably well despite poor composition estimates. In contrast, real data involve time-varying scrap composition and uncertainties, making  it more challenging for windowed NNLS to perform effectively.

Figure~\ref{fig:realdata_KF_cu2_f}, Figure~\ref{fig:realdata_KF_cu36_f} and Figure~\ref{fig:realdata_KF_cu37_f} present the estimated Cu fraction for selected scrap types. For NNLS, initial estimates are unreliable due to insufficient data and are not shown in the figures.
The figures reveal that the results of the windowed NNLS fluctuate significantly, which is unrealistic for practical applications. Although it is not listed in the results,  as the window size increases from $2000$ to the number of total heats, the fluctuations diminish,  and the results tent to converge to a constant, which is also impractical. In contrast, the Kalman filter results remain within a plausible and realistic range.

\begin{table}[h!]
    \centering
    \setlength{\tabcolsep}{10pt}
    \caption{Comparison of mean and standard deviation of error($f_{\rm steel ,Cu}$) for NNLS and Kalman filter methods (unit: ppm).}
    \label{tab:ols_kalman_comparison}
    \begin{tabular}{SSS}
        \toprule
        {Method}          & {mean (error)}  & {std (error)} \\
        \midrule
        {Windowed NNLS}   & {$-63$}         & {$67$}  \\
        {Kalman filter}   & {$-11$}         & {$52$}  \\
        \bottomrule
    \end{tabular}
\end{table}

\subsection{Non-linear model -- Cr}
Based on equations~\eqref{eq:mass_balance_crs_f}--\eqref{eq:mass_balance_crs}, the model for Cr can be treated as a linear model if $\ell_{\rm Cr}$ is regarded as a constant
\[
\sum_{s=1}^{N_{\rm s}} m_s \, \alpha_{s, {\rm Cr}} = 
f_{\text{steel},\, \rm{Cr}} \,(m_{\text{steel}} + m_{\text{slag}}\, \ell_{\rm Cr})
- m_{\text{hm}}\, f_{\text{hm}, \,\rm {Cr}},
\]
where the Cr fraction in the scrap type can be solved using windowed NNLS, too. Here we  choose the window size to be $2000$ heats for the same reason as before.

For the unscented Kalman filter, the mean vector $\vec{q}$ for the Cr fractions in the scrap types is also solved by NNLS using the first $5000$ heats. The mean vector $\vec{q}_c$ for the partition coefficients is manually selected to be $\vec{q}_c = [ 9.7, \,  0.01]^\top$, based on expert knowledge. The initial mean vector is set to $\vec a_1^+ = [\vec q, \, \vec q_c ]^\top$. The  matrix $P_\infty^+$ is chosen diagonal with elements $(P_\infty^+)_{i,i} = (0.05 (\vec a^+_1)_i)^2$, where $(\vec a_1^+)_i$ is the $i$-th component of vector $\vec a_1^+$. The parameter $\gamma$ is set to $7 \times 10^{-6}$, and the observation noise variance is set to  $H_t = 330^2 \times 4^2$.
For the same reason,  predictions are evaluated against measured values for $f_{\rm steel, \,Cr}$, using historical data exclusively (excluding predicted heats). The first $10000$ heats are not used for calculating the error. 

Table~\ref{tab:ols_ukf_comparison} and Figure~\ref{fig:cr_realdata_output} show the error of predicting $f_{\rm steel, Cr}$. As can be seen, the UKF results exhibit lower bias and standard deviation, indicating improved accuracy, compared to NNLS.
Figure~\ref{fig:realdata_KF_cr2_f}, Figure~\ref{fig:realdata_KF_cr36_f} and Figure~\ref{fig:realdata_KF_cr37_f} present the estimated Cr fraction for selected scrap types. For NNLS, initial estimates are unreliable due to insufficient data and are not shown in the figures. These results are consistent with the results from previous subsections. These results suggest that the UKF significantly outperforms windowed NNLS for non-linear models in this context.

\begin{table}[h!]
    \centering
    \setlength{\tabcolsep}{10pt}
    \caption{Comparison of mean and standard deviation of error($f_{\rm steel ,Cr}$) for NNLS and Kalman filter method (unit: ppm).}
    \label{tab:ols_ukf_comparison}
    \begin{tabular}{SSS}
        \toprule
        {Method}          & {mean (error)} & {std (error)} \\
        \midrule
        {Windowed NNLS}   & {$-56$}        & {$63$} \\
        {UKF}             & {$-13$}        & {$55$} \\
        \bottomrule
    \end{tabular}
\end{table}

\section{Conclusion and discussion}\label{s:conclusion}

This study presented linear and non-linear state space models for estimating the composition of scrap in Electric Arc Furnace (EAF) and Basic Oxygen Furnace (BOF) steelmaking. To fit these models, a modified Kalman filter and a modified unscented Kalman Filter (UKF) were developed for the linear and non-linear cases, respectively, along with practical suggestions for hyperparameter selection.  Cu and Cr are chosen as the representative elements for linear and non-linear models, and tested on real production data.

The windowed NNLS method was chosen as a simple benchmark for comparison, as it has a clear physical interpretation aligned with the steelmaking context and is straightforward to implement. The results showed that the Kalman filter and UKF provided more accurate scrap composition estimates, outperforming windowed non-negative least squares (NNLS) regression. Both filters demonstrated lower bias and standard deviation in error measurements compared to NNLS.

The models effectively captured the time-variant nature of scrap composition, making them suitable for real-time steel production control. However, some limitations were identified, such as the requirement for  hyperparameter tuning and the assumption of sufficiently accurate measurements for variables like input scrap mass.

Future work could focus on enhancing the models by incorporating automated hyperparameter tuning techniques to improve robustness and applicability in industrial settings. Additionally, a more comprehensive model considering measurement noise across all variables could be proposed. It may also be valuable to integrate this approach with models that consider gases behavior in steelmaking processes.

\bibliographystyle{abbrv} 
\bibliography{reference}

\end{document}